\documentclass[aps,preprint,showpacs,amsmath,amssymb,eqsecnum,eqnarray,array]{revtex4}
\usepackage{graphicx}
\usepackage{dcolumn}
\usepackage{bm}
\begin{document}

\title{Theory of confined plasmonic waves in coaxial cylindrical cables fabricated of metamaterials}
\author{M. S. Kushwaha$^{1}$ and B. Djafari-Rouhani$^{2}$}
\address
{$^{1}$Department of Physics and Astronomy, Rice University, P.O. Box 1892, Houston, TX 77251, USA\\
 $^{2}$IEMN, UMR-CNRS 8520, UFR de Physique, University of Science and Technology of Lille I, 59655 Villeneuve
 d'Ascq Cedex, France}

\date{\today}

\begin{abstract}
We report on the theoretical investigation of the plasmonic wave propagation in the coaxial cylindrical
cables fabricated of both right-handed medium (RHM) [with $\epsilon >0$, $\mu >0$] and left-handed
medium (LHM) [with $\epsilon(\omega) <0$, $\mu(\omega) <0$], using a Green-function (or response function)
theory in the absence of an applied magnetic field. The Green-function theory generalized to be
applicable to such quasi-one dimensional systems enables us to derive explicit expressions
for the corresponding response functions (associated with the EM fields), which can in turn be
used to derive various physical properties of the system. The confined plasmonic wave excitations in
such multi-interface structures are characterized by the electromagnetic fields that are
localized at and decay exponentially away from the interfaces.  A rigorous analytical diagnosis of
the general results in diverse situations leads us to reproduce exactly the previously well-known
results in other geometries, obtained within the different theoretical frameworks. As an
application, we present several illustrative examples on the dispersion characteristics of the
confined (and extended) plasmonic waves in single- and double-interface structures made up of
dispersive metamaterials interlaced with conventional dielectrics. These dispersive modes are also
substantiated through the computation of local as well as total density of states. It is observed
that the dispersive components enable the system to support the simultaneous existence of s- and
p-polarization modes in the system. Such effects as this one are solely attributed to the
negative-index metamaterials and are otherwise impossible. The readers will also notice the explicit
$\mu$-dependence of the dispersion relations for the s-polarization modes, obtained under special
limits in some cases, for the single- and double-interface systems. The elegance of the theory lies in
the fact that it does not require the matching of the boundary conditions and in its simplicity and
the compact form of the desired (analytical) results.
\end{abstract}
\maketitle

\newpage

\section{INTRODUCTION}

The discovery of quantum Hall effects spurred a tremendous research interest in the quantum phenomena associated with the systems of lower dimensions such as quantum wells, quantum wires, and quantum dots. While this momentum still seems to be growing [1], the classical phenomena emerged with the proposal of the photonic (and phononic) crystals and, more recently, the negative-index metamaterials have been drawing considerable attention of many research groups world-wide. Proposed some four decades ago by Veselago [2], advocated by Sir John Pendry in 2000 [3], and first practically realized by Smith and coworkers in 2001 [4], an artificially designed negative-index metamaterial, exhibiting simultaneously negative values of electrical permittivity $\epsilon(\omega)$ and magnetic permeability $\mu(\omega)$
and hence negative refractive index $n$, seems to have changed many basic notions related with the electromagnetism. It forms a left-handed medium, with the energy flow ${\bf E\times {\bf H}}$ being opposite to the direction of propagation, for which it has been argued that such phenomena as Snell's law, Doppler effect, Cherenkov radiation are inverted.

The subject of composite systems made up of metamaterials has gained un unexpected momentum and the research interest
seems to have focused not only on the photonic crystals with metamaterial components [5-12] but also on the single- and multi-layered planar structures [13-22] as well as on the (usually) single cylindrical geometries [23-34]. The interesting phenomena emerging from the geometries involving metamaterials include the slowing, trapping, and releasing of the light signals [35], the proposal of the cloaking devices [36], and the extraordinary refraction of light [37]. The early development of the subject can be found in interesting review articles by Pendry [38] and by Boardman [39]. Cloaking is an illusion like a mirage: you steer light around an object and therefore you never see the object.
The tailored response of the metamaterials has had a dramatic impact on engineering, optics, and physics communities alike, because they can offer electromagnetic properties that are difficult or impossible to achieve with naturally occurring materials.

The recent research interest in surface plasmonic wave optics has been invigorated by the experiment performed on the transmission of light through subwavelength holes in metal films [40]. This experiment has spurred numerous theoretical [41-44] as well experimental [45-49] works on similar structured surfaces: either perforated with holes, slits, dimples, or decorated with grooves. It has been argued that resonant excitation of surface plasmonic waves creates huge electric fields at the surface that force the light through the holes, yielding very high transmission coefficients. The idea of tailoring the topography of a perfect conductor to support the surface waves resembling the behavior of the surface plasmonic waves at optical frequencies was discussed in the context of a surface with an array of two-dimensional holes [44]. The experimental verification of this proposal has recently been reported [50-52] on the structured metamaterial surfaces which support surface plasmonic waves at microwave frequencies.

The purpose of this paper is to investigate the plasmonic wave propagation in the coaxial cables fabricated of metamaterials interlaced with conventional dielectrics using the Green-function (or response function) theory in the absence of an applied magnetic field. The roots of our theoretical approach lie virtually in the interface-response theory (IRT) [53] generalized to be applicable to such quasi-one dimensional (Q1D) systems. Ever since its inception, the IRT has been extensively applied to study various quasi-particle excitations (such as phonons, plasmons, magnons, etc.) in heterostructures and superlattices [54-56].

The use of IRT has numerous advantages over the traditional Maxwell equations with boundary conditions. It is very well known that the Maxwell equations with proper boundary conditions only provide us with the dispersion relations for the
given electromagnetic waves in an inhomogeneous medium. The IRT, which is essentially a matrix formulation, on the other hand, enables us to obtain not only the dispersion relations of the desired excitations but also provides us with a wide platform to study various static and dynamic properties in terms of the response functions of the resultant systems at hand. These include, for instance, the local and total density of states, reflection and transmission coefficients,
inelastic light and electron scattering, tunneling phenomena, and selective transmission, to name a few.


The rest of the paper is organized as follows. In Sec. II we discuss some relevant basic notions of the cylindrical geometry and calculate the bulk response function. In Sec. III we present theoretical formalism to derive the final expressions for the plasmonic wave dispersion relations in the compact form, discuss some analytical diagnoses of the
general expressions under special limits, and give some explicit relationship between the response functions and
the density of states. In Sec. IV we report several interesting illustrative examples on the plasmonic wave dispersion and density of states in a variety of experimentally feasible situations. Finally, we conclude our findings and list
some interesting dimensions worth adding to the problem in the future in Sec. V.

\section{BASIC NOTIONS AND BULK RESPONSE FUNCTIONS}

We consider the electromagnetic waves propagating with an angular frequency $\omega$ and wave vector
$k\parallel \hat{z}$ in a medium defined by the cylindrical coordinates $(\rho,\theta, z)$. The
plasmonic waves, here as well as in the later part of this work, will be assumed to observe the spatial
localization along the direction perpendicular to the axis of the cylinder. Note that the situation
is totally unlike that in the Cartesian co-ordinate system where one can safely and readily define
the sagittal plane (i.e., the plane defined by the wave vector and the normal to the surface/interface)
and hence isolate the transverse magnetic (TM) and the transverse electric (TE) modes, at least in the
absence of an applied magnetic field. The only exception to this notion is the Voigt geometry (with a
magnetic field parallel to the surface/interface and perpendicular to the propagation vector) that
can still (i.e., even in the presence of an applied magnetic field) allow the separation of the TM and
TE modes (see, for details, Ref. 1). In the literature on optics the TM and TE modes are also known by
the name of p-polarization and s-polarization, respectively.

It should be pointed out that since we are interested in the artificially designed metamaterials whose response is measured both in terms of the effective $\epsilon$ and $\mu$, $\vec{B}\neq \vec{H}$ in the Maxwell curl-field
equations. After eliminating the magnetic field variable $\vec{B}$ from these curl-field equations, we obtain

\begin{equation}
\vec{\nabla}\times (\vec{\nabla}\times \vec{E}) - q_{_{0}}^2\,\epsilon\,\mu\, \vec{E}=0 \\
\end{equation}
Here both $\epsilon$ and $\mu$ are  scalar quantities, since the system we are concerned with is not subjected to
any external magnetic field and the physical system is assumed to be isotropic. In Eq. (2.1) $q_{_{0}}=\omega/c$ is
the vacuum wave vector, where $c$ is the speed of light in vacuum. We will take the spatial and temporal dependence
of the electromagnetic fields to be of the form of
$\vec{A}(\rho,\theta,z)\,\sim \,\vec{A}(\rho, \theta)\,\,e^{(i k z-i\omega t)}$,
where $\vec{A}\equiv \vec{E}$ or $\vec{B}$. Recalling the standard definitions of
$\vec{\nabla}.\vec{A}$, $\nabla^2\phi$ (with $\phi$ as any scalar), and $\vec{\nabla}\times \vec{A}$
in the cylindrical coordinates, one should be able to split Eq. $(2.1)$ in the three equations:

\begin{equation}
\left[\frac{\partial^2}{\partial \rho^2}+\frac{1}{\rho}\frac{\partial}{\partial \rho}+
\frac{1}{\rho^2}\frac{\partial ^2}{\partial \theta^2}-k^2 \right]E_{\rho} -
\frac{1}{\rho^2}\left(E_{\rho}+2\frac{\partial}{\partial \theta}E_{\theta}\right)+
q_{_{0}}^2\,\epsilon\,\mu\, E_{\rho}=0\\
\end{equation}
\begin{equation}
\left[\frac{\partial^2}{\partial \rho^2}+\frac{1}{\rho}\frac{\partial}{\partial \rho}+
\frac{1}{\rho^2}\frac{\partial^2}{\partial \theta^2}-k^2 \right]E_{\theta} -
\frac{1}{\rho^2}\left(E_{\theta}-2\frac{\partial}{\partial \theta}E_{\rho}\right)+
q_{_{0}}^2\,\epsilon\,\mu\, E_{\theta}=0\\
\end{equation}
\begin{equation}
\left[\frac{\partial^2}{\partial \rho^2}+\frac{1}{\rho}\frac{\partial}{\partial \rho}+
\frac{1}{\rho^2}\frac{\partial^2}{\partial \theta^2}-k^2 \right]E_z +
q_{_{0}}^2\,\epsilon\,\mu\, E_z=0\\
\end{equation}
Equations $(2.2)-(2.4)$ demonstrate it clearly that the cylindrical geometry prevents the separation
of the TM and TE modes. We choose to work in terms of $E_z$ and $H_z$ components. Then we first need
to evaluate $E_{\rho}$, $E_{\theta}$, $H_{\rho}$, and $H_{\theta}$ in terms of $E_z$ and $H_z$ from
the Maxwell curl-field equations. The result is

\begin{equation}
E_{\rho}=\frac{1}{\alpha^2}\left [-iq_{_{0}}{\mu}\frac{1}{\rho}\frac{\partial}{\partial \theta}H_z
- ik\frac{\partial}{\partial \rho}E_z \right ]\\
\end{equation}
\begin{equation}
E_{\theta}=\frac{1}{\alpha^2}\left [ iq_{_{0}}{\mu}\frac{\partial}{\partial \rho}H_z -
ik\frac{1}{\rho}\frac{\partial}{\partial \theta}E_z \right ]\\
\end{equation}
and similarly

\begin{equation}
H_{\rho}=\frac{1}{\alpha^2}\left [iq_{_{0}}\,\epsilon\,\frac{1}{\rho}\frac{\partial}{\partial
\theta}E_z - ik\frac{\partial}{\partial \rho}H_z\right ]\\
\end{equation}
\begin{equation}
H_{\theta}=\frac{1}{\alpha^2}\left [-iq_{_{0}}\,\epsilon\,\frac{\partial}{\partial \rho}E_z -
ik\frac{1}{\rho}\frac{\partial}{\partial \theta}H_z\right ]\\
\end{equation}
With the aid of these equations, we simplify the z-components of the Maxwell curl-field equations:

\begin{equation}
-iq_{_{0}}\,\epsilon\,E_z=\frac{1}{\rho}\frac{\partial}{\partial \rho}(\rho H_{\theta})-
\frac{1}{\rho}\frac{\partial}{\partial \theta}H_{\rho}
\end{equation}
and

\begin{equation}
iq_{_{0}}\,\mu\,H_z=\frac{1}{\rho}\frac{\partial}{\partial \rho}(\rho E_{\theta})-
\frac{1}{\rho}\frac{\partial}{\partial \theta}E_{\rho}
\end{equation}
to write

\begin{equation}
\frac{\partial^2}{\partial \rho^2}A_z + \frac{1}{\rho}\frac{\partial}{\partial \rho}A_z +
\left(\frac{1}{\rho^2}\frac{\partial^2}{\partial \theta^2}-\alpha^2\right)A_z=0 \\
\end{equation}
where $A_z$ stands for $E_z$ or $H_z$ and $\alpha=(k^2-q_{_{0}}^2\,\epsilon\,\mu)^{1/2}$ is a measure
of the decay constant in a medium concerned.

Before proceeding further, it is important to define a characteristic terminology of the interface response theory:
the black-box surface (BBS). By BBS we mean an entirely opaque surface through which electromagnetic fields cannot propagate. The idea of introducing the BBS in the IRT [53] was conceived with two prominent advantages over the contemporary semiclassical approaches in mind. Firstly, it allows one to disconnect completely from the extra mathematical world and hence to confine stringently within the truly building block of the system concerned.
Secondly, it implicitly provides a great opportunity to get rid of using the boundary conditions one is so routinely
accustomed to in dealing with the inhomogeneous systems. What results is a number of simplified and compact forms of
the response functions which one only needs to sum up in order to proceed further for studying the desired physical property of the resultant system at hand. Conceptually, this is achieved by imposing that $c$ (the speed of light), $\epsilon$ (the electric permittivity), and  $\mu$ (the magnetic permeability) vanish inside the specific region.
This region represents a medium that could, in principle, be semi-infinite or finite. In order to create a medium bounded by a black-box surface, we assume that the Eqs. $(2.5)-(2.8)$ are only valid for either $\rho > R$ or
$\rho < R$, with $R$ as the radius of the only cylinder in question by now. Then we multiply the right-hand sides of Eqs. $(2.5)-(2.8)$ by the step function $\theta(\rho-R)$ or $\theta(R-\rho)$, as the case may be. We first calculate
the two derivatives needed to evaluate Eqs. (2.9) and (2.10). The result is

\begin{eqnarray}
\frac{\partial}{\partial \rho}(\rho E_{\theta})=\frac{1}{\alpha^2}\left \{
\left[iq_{_{0}}\mu \frac{\partial}{\partial \rho}H_z +
iq_{_{0}}\mu \rho\frac{\partial^2}{\partial \rho^2}H_z -
ik\frac{\partial}{\partial \theta}\frac{\partial}{\partial \rho}E_z\right] \right.\nonumber\\
-\left.\delta (R-\rho)\left[iq_{_{0}}\mu \rho\frac{\partial}{\partial \rho}H_z -
ik\frac{\partial}{\partial \theta}E_z\right] \right\}
\end{eqnarray}
and

\begin{eqnarray}
\frac{\partial}{\partial \rho}(\rho H_{\theta})=\frac{1}{\alpha^2}\left \{
\left[-iq_{_{0}}\,\epsilon\,\frac{\partial}{\partial \rho}E_z +
iq_{_{0}}\,\epsilon\,\rho\frac{\partial^2}{\partial \rho^2}E_z -
ik\frac{\partial}{\partial \theta}\frac{\partial}{\partial \rho}H_z\right]\right. \nonumber\\
-\left.\delta (R-\rho)\left[-iq_{_{0}}\,\epsilon\,\rho\frac{\partial}{\partial \rho}E_z -
ik\frac{\partial}{\partial \theta}H_z\right] \right\}
\end{eqnarray}
Evidently, the step function (and hence the delta function) dictates the kind of physical situation
we will consider in what follows.  Then the differential equations (2.9) and (2.10) satisfied by
$E_z(\rho, \theta)$ and $H_z(\rho, \theta)$ assume the following forms:

\begin{equation}
\left(\frac{-iq_{_{0}}\,\epsilon}{\beta^2} \right) \left[
\left(\frac{\partial^2}{\partial\rho^2} +
\frac{1}{\rho}\frac{\partial}{\partial \rho} +
\frac{1}{\rho^2}\frac{\partial^2}{\partial \theta^2} + \beta^2 \right)E_z -
\delta (R-\rho)\left(\frac{\partial}{\partial \rho}E_z  +
\frac{k}{q_{_{0}}\,\epsilon\,\rho}\frac{\partial}{\partial \theta}H_z\right)\right]=0
\end{equation}
and

\begin{equation}
\left(\frac{iq_{_{0}}\,\mu}{\beta^2} \right) \left[
\left(\frac{\partial^2}{\partial\rho^2} +
\frac{1}{\rho}\frac{\partial}{\partial \rho} +
\frac{1}{\rho^2}\frac{\partial^2}{\partial \theta^2} + \beta^2 \right)H_z -
\delta (R-\rho)\left(\frac{\partial}{\partial \rho}H_z  -
\frac{k}{q_{_{0}}\,\mu\,\rho}\frac{\partial}{\partial \theta}E_z\right)\right]=0
\end{equation}
where $\beta^2=-\alpha^2=q_{_{0}}^2\epsilon\,\mu - k^2$. The formal equations (2.14) and (2.15) will
be the standard format for all the calculations of the Green functions of the system of interest in
what follows.

Next, let $\vec{r}\equiv (\rho, \theta)$,  $\vec{r}'\equiv (\rho', \theta')$, and define the Green
function

\begin{equation}
G\left(\vec{r},\vec{r}'\right)\equiv G\left( \mid \vec{r} - \vec{r}' \mid \right)\equiv
G\left( \rho,\theta; \rho', \theta'\right) 
\end{equation}
for the homogeneous (bulk) medium [see Eq. (2.11)]

\begin{eqnarray}
\left( \frac{\partial^2}{\partial \rho^2} + \frac{1}{\rho}\frac{\partial}{\partial \rho} +
\frac{1}{\rho^2}\frac{\partial^2}{\partial \theta^2} + \beta^2 \right)G(\vec{r}, \vec{r}')
  =  -4\pi \delta (\vec{r}-\vec{r}') \nonumber \\
  =  -\frac{4\pi}{\rho}\delta(\rho-\rho')\delta(\theta-\theta')
\end{eqnarray}
The solution of this equation is given by (see, for example, Ref. [57]):

\begin{equation}
G(\vec{r}, \vec{r}')=\sum_{m=-\infty}^{\infty} e^{im(\theta-\theta')}\,G(m; \rho,\rho')
\end{equation}
with

\begin{eqnarray}
G(m;\rho,\rho')=i\pi\left\{
\begin{array}{rrl}
J_m(\beta \rho)H_m(\beta \rho') & ,\ \ \ \ \ & \mbox{if}\ \ \ \  \rho \leq \rho'\\
H_m(\beta \rho)J_m(\beta \rho') & ,\ \ \ \ \ & \mbox{if} \ \ \ \ \rho \geq \rho'
\end{array}
\right.
\end{eqnarray}
where $J_m(z)$ [$H_m(z)$] refers to the Bessel function of the first (third) kind of (integer) order
$m$. We write

\begin{eqnarray}
G(m;\rho,\rho')=i\pi \left\{ [1  -  \theta(\rho-\rho')]J_m(\beta \rho)H_m(\beta \rho')\right. \nonumber \\
   +  \left.\theta (\rho-\rho')H_m(\beta \rho)J_m(\beta \rho') \right\}
\end{eqnarray}
where $\theta(x)=1(0)$ for $x>0 (x<0)$ is the Heaviside step function. It is not difficult to verify
that the Green function in Eq. (2.19) represents the exact solution of Eq. (2.18).

We close this section by writing the bulk Green-function tensor for the field components $E_z$ and $H_z$
as a $2\times 2$ matrix:

\begin{eqnarray}
\left[
\begin{array}{cc}
\left(\frac{-q_{_{0}}^2\,\epsilon}{\beta^2}\right)\left[\frac{\partial^2}{\partial \rho^2}+
\frac{1}{\rho}\frac{\partial}{\partial \rho}-\frac{m^2}{\rho^2}+\beta^2\right] & 0 \\
0 & \left(\frac{-q_{_{0}}^2\,\mu}{\beta^2}\right)\left[\frac{\partial^2}{\partial \rho^2}+
\frac{1}{\rho}\frac{\partial}{\partial \rho}-\frac{m^2}{\rho^2}+\beta^2\right]
\end{array}
\right] \nonumber \\
\nonumber\\
\times\,\left[
\begin{array}{cc}
G_{_{E}}(m;\rho,\rho') & 0 \\
0 & G_{_{H}}(m;\rho,\rho')
\end{array}
\right]
=-\,\frac{2}{\rho}\,\,\delta(\rho-\rho')
\left[
\begin{array}{cc}
1 & 0\\
0 & 1
\end{array}
\right]
\end{eqnarray}
where we use $\partial/\partial \theta=im$ and [see Eq. (2.19)]

\begin{eqnarray}
-\left(\frac{q_{_{0}}^2\,\epsilon}{\beta^2}\right)G_{_{E}}(m;\rho,\rho')
=&&-\,\left(\frac{q_{_{0}}^2\,\mu}{\beta^2}\right)G_{_{H}}(m;\rho,\rho') \nonumber \\
=&&i\pi\left\{
\begin{array}{lll}
J_m(\beta \rho)H_m(\beta \rho')\ \ \ & , & \ \ \ \rho \leq \rho' \\
H_m(\beta \rho)J_m(\beta \rho')\ \ \ & , & \ \ \ \rho \geq \rho'
\end{array}
\right.
\end{eqnarray}
In what follows, we will consider three types of perturbative operations to have the desired results for the resultant structure at hand.  In doing so, we will abide by the conceptual scheme of the IRT (see Ref. [53]).

\section{FORMALISM FOR INVERSE RESPONSE FUNCTIONS}

In this section, we will consider three perturbative operations represented geometrically by Fig. 1. Specifically, Fig. 1(A), 1(B), and 1(C) correspond, respectively, to the metamaterial cylinder of radius $R_1$ surrounded by a black box surface, a black box cylinder of radius $R_2$ surrounded by a metamaterial, and a metamaterial shell sandwiched between the black box cylinder of radius $R_1$ and a semi-infinite black box surface outside a cylinder of radius $R_2$. The metamaterial media in the perturbations A, B, and C are, in general, characterized by the local permittivity and permeability functions $\epsilon_1(\omega)$, $\mu_1(\omega)$; $\epsilon_2(\omega)$, $\mu_2(\omega)$;  and $\epsilon_3(\omega)$, $\mu_3(\omega)$ respectively. We will consider the effect of retardation but neglect the absorption throughout. Any subscript $i\equiv 1$, 2, or 3 on the physical quantities should be understood referring to the respective perturbation until and unless stated otherwise.

\begin{figure}[htbp]
\includegraphics*[width=9cm,height=10cm]{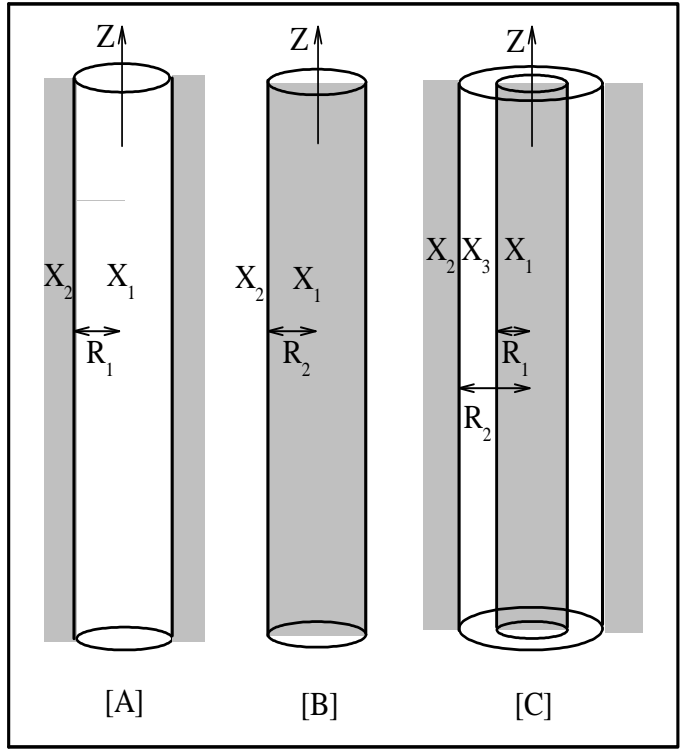}
\caption{Schematics of the concept of three perturbations: [A], [B], and [C]. The blank (shaded) region refers to the material medium (black box) in the system. The sum of the first two perturbations defines
a plasma (dielectric) cylinder embedded in a dielectric (plasma) and the sum of all three perturbations specifies, say, 
a plasma (dielectric) shell surrounded by two unidentical dielectrics (plasmas). Here $R_j$ is the radius and $X\equiv \epsilon(\omega)$ and/or $\mu(\omega)$ for a specified medium.}
\label{fig1}
\end{figure}

\subsection{First Perturbation}

The first perturbation [represented by Fig. 1(A)] is specified by a step function $\theta (R_1-\rho)$ in front of Eqs. $(2.5) - (2.8)$. That means that the black-box cleavage operator $\tilde{V}_1(R_1,\rho')\delta(R_1-\rho')$ is defined such that (see Eqs. (2.14)-(2.15))

\begin{equation}
\tilde{V}_1(R_1,\rho')=\frac{R_1}{2}\,\frac{q_{_{0}}^2}{\beta_1^2}
\left[
\begin{array}{cc}
-\epsilon_1 \frac{\partial}{\partial \rho'}\  &  \  -\frac{imk}{q_{_{0}}\rho'} \\
\frac{imk}{q_{_{0}}\rho'} \  & \ -\mu_1\frac{\partial}{\partial \rho'}
\end{array}
\right]
\end{equation}
and the corresponding bulk Green function is written as [see Eqs. (2.22)]

\begin{equation}
\tilde{G}_1(\rho,\rho')=i\pi\, \frac{\beta_1^2}{q_{_{0}}^2}
\left[
\begin{array}{cc}
-\frac{1}{\epsilon_1}H_m(\beta_1\rho)J_m(\beta_1\rho') \  & \ 0 \\
0 \ & \  -\frac{1}{\mu_1} H_m(\beta_1\rho)J_m(\beta_1\rho')
\end{array}
\right]
\end{equation}
It is noteworthy that although the operators $\tilde{V}_1$ and $\tilde{A}_1$ as well as the functions $\tilde{G}_1$
and $\tilde{g}_1$ are all functions of the variables such as $m$, $k$, and $\omega$, we have suppressed them
throughout for the sake of brevity and convenience. With this, we define the response operator

\begin{eqnarray}
\tilde{A}_1\left(R_1,R_1\right)
=&&\tilde{V}_1(R_1,\rho)\,\tilde{G}_1(\rho,\rho')\mid_{\rho=R_1=\rho'}\nonumber\\
\nonumber \\
=&&\left[
\begin{array}{cc}
\frac{i\pi}{2}\beta_1R_1 H'_m(\beta_1R_1)J_m(\beta_1R_1) \ & \
-\frac{\pi}{2}\frac{mk}{q_{_{0}}\mu_1}H_m(\beta_1R_1)J_m(\beta_1R_1) \\
\frac{\pi}{2}\frac{mk}{q_{_{0}}\epsilon_1}H_m(\beta_1R_1)J_m(\beta_1R_1) \ & \
\frac{i\pi}{2}\beta_1R_1H'_m(\beta_1R_1)J_m(\beta_1R_1)
\end{array}
\right]
\end{eqnarray}
The prime on the Bessel functions stands for the derivative of the respective quantity with respect to
the full argument. Next we define an operator

\begin{eqnarray}
\tilde{\Delta}_1(R_1,R_1)=&&
\tilde{I}\,+\,\tilde{A}_1(R_1,R_1) \nonumber \\
\nonumber \\
=&&\left[
\begin{array}{cc}
\frac{i\pi}{2}\beta_1R_1 H_m(\beta_1R_1)J'_m(\beta_1R_1) \ & \
-\frac{\pi}{2}\frac{mk}{q_{_{0}}\mu_1}H_m(\beta_1R_1)J_m(\beta_1R_1) \\
\frac{\pi}{2}\frac{mk}{q_{_{0}}\epsilon_1}H_m(\beta_1R_1)J_m(\beta_1R_1) \ & \
\frac{i\pi}{2}\beta_1R_1H_m(\beta_1R_1)J'_m(\beta_1R_1)
\end{array}
\right]
\end{eqnarray}
It should be pointed out that in writing the second equality in Eq. (3.4), we have made use of the
identity [58]

\begin{equation}
\frac{1}{J_{\nu}(z)H_{\nu}(z)}=\frac{\pi z}{2i}
\left[\frac{H'_{\nu}(z)}{H_{\nu}(z)}-\frac{J'_{\nu}(z)}{J_{\nu}(z)}\right]
\end{equation}
Next, we calculate the inverse of $\tilde{G}_1$ to write

\begin{equation}
\tilde{G}^{-1}_1(R_1,R_1)=\frac{q_{_{0}}^2}{i\pi\beta_1^2}\,\frac{1}{H_m(\beta_1R_1)J_m(\beta_1R_1)}
\left[
\begin{array}{cc}
-\epsilon_1  \ & \ 0  \\
0 \ & \  -\mu_1
\end{array}
\right]
\end{equation}
As such, we have all what we need to calculate the inverse response function in the interface space $M$
defined by

\begin{equation}
\tilde{g}^{-1}_1(R_1,R_1)= \tilde{\Delta}_1(R_1, R_1)\,\tilde{G}^{-1}_1(R_1, R_1)
\end{equation}
The result is that

\begin{equation}
\tilde{g}^{-1}_1(R_1,R_1)=\frac{q_{_{0}}^2}{2\beta_1^2}
\left[
\begin{array}{cc}
- \beta_1R_1\epsilon_1 \frac{J'_m(\beta_1R_1)}{J_m(\beta_1R_1)} \ \ & \ \ -\frac{imk}{q_{_{0}}}\\
\frac{imk}{q_{_{0}}} \ \ & \ \  -\beta_1R_1\mu_1 \frac{J'_m(\beta_1R_1)}{J_m(\beta_1R_1)}
\end{array}
\right]
\end{equation}
represents the response function of a material medium cylinder surrounded by a black box. By material medium we mean
a medium possessing a realistic substance such as a metal, dielectric, semiconductor, or metamaterial.

\subsection{Second Perturbation}

The second perturbation [represented by Fig. 1(B)] is specified by a step function $\theta (\rho-R_2)$ in front of
Eqs. $(2.5) - (2.8)$. Then the black-box cleavage operator $\tilde{V}_2(R_2,\rho')\delta(\rho'-R_2)$ is defined
such that

\begin{equation}
\tilde{V}_2(R_2,\rho')=-\frac{R_2}{2}\,\frac{q_{_{0}}^2}{\beta_2^2}
\left[
\begin{array}{cc}
-\epsilon_2 \frac{\partial}{\partial \rho'}\  &  \  -\frac{imk}{q_{_{0}}\rho'} \\
\frac{imk}{q_{_{0}}\rho'} \  & \ -\mu_2\frac{\partial}{\partial \rho'}
\end{array}
\right]
\end{equation}
and the corresponding bulk Green function is written as

\begin{equation}
\tilde{G}_2(\rho,\rho')=i\pi\, \frac{\beta_2^2}{q_{_{0}}^2}
\left[
\begin{array}{cc}
-\frac{1}{\epsilon_2}J_m(\beta_2\rho)H_m(\beta_2\rho') \  & \ 0 \\
0 \ & \ - \frac{1}{\mu_2}J_m(\beta_2\rho)H_m(\beta_2\rho')
\end{array}
\right]
\end{equation}
With this, we define the response operator

\begin{eqnarray}
\tilde{A}_2\left(R_2,R_2\right)
=&&\tilde{V}_2(R_2,\rho)\,\tilde{G}_2(\rho,\rho')\mid_{\rho=R_2=\rho'}\nonumber\\
\nonumber \\
=&&\left[
\begin{array}{cc}
-\frac{i\pi}{2}\beta_2R_2 J'_m(\beta_2R_2)H_m(\beta_2R_2) \ & \
+\frac{\pi}{2}\frac{mk}{q_{_{0}}\mu_2}J_m(\beta_2R_2)H_m(\beta_2R_2) \\
-\frac{\pi}{2}\frac{mk}{q_{_{0}}\epsilon_2}J_m(\beta_2R_2)H_m(\beta_2R_2) \ & \
-\frac{i\pi}{2}\beta_2R_2J'_m(\beta_2R_2)H_m(\beta_2R_2)
\end{array}
\right]
\end{eqnarray}
Next, we define an operator

\begin{eqnarray}
\tilde{\Delta}_2(R_2,R_2)=&&
\tilde{I}\,+\,\tilde{A}_2(R_2,R_2) \nonumber \\
\nonumber \\
=&&\left[
\begin{array}{cc}
-\frac{i\pi}{2}\beta_2R_2 J_m(\beta_2R_2)H'_m(\beta_2R_2) \ & \
+\frac{\pi}{2}\frac{mk}{q_{_{0}}\mu_2}J_m(\beta_2R_2)H_m(\beta_2R_2) \\
-\frac{\pi}{2}\frac{mk}{q_{_{0}}\epsilon_2}J_m(\beta_2R_2)H_m(\beta_2R_2) \ & \
-\frac{i\pi}{2}\beta_2R_2J_m(\beta_2R_2)H'_m(\beta_2R_2)
\end{array}
\right]
\end{eqnarray}
Again, in writing the second equality in Eq. (3.12), we have made use of the identity in Eq. (3.5).
Next, we calculate the inverse of $\tilde{G}_2$ to write

\begin{equation}
\tilde{G}^{-1}_2(R_2,R_2)=\frac{q_{_{0}}^2}{i\pi\beta_2^2}\,\frac{1}{J_m(\beta_2R_2)H_m(\beta_2R_2)}
\left[
\begin{array}{cc}
-\epsilon_2  \ & \ 0  \\
 0 \ & \  -\mu_2
\end{array}
\right]
\end{equation}
Now we need to calculate the inverse response function in the interface space $M$ defined by

\begin{equation}
\tilde{g}^{-1}_2(R_2,R_2)= \tilde{\Delta}_2(R_2, R_2)\,\tilde{G}^{-1}_2(R_2, R_2)
\end{equation}
The result is that

\begin{equation}
\tilde{g}^{-1}_2(R_2,R_2)=\frac{q_{_{0}}^2}{2\beta_2^2}
\left[
\begin{array}{cc}
 \beta_2R_2\epsilon_2 \frac{H'_m(\beta_2R_2)}{H_m(\beta_2R_2)} \ \ & \ \ \frac{imk}{q_{_{0}}}\\
-\frac{imk}{q_{_{0}}} \ \ & \ \  \beta_2R_2\mu_2 \frac{H'_m(\beta_2R_2)}{H_m(\beta_2R_2)}
\end{array}
\right]
\end{equation}
represents the response function of a black box surrounded by a material medium.

\subsection{Third Perturbation}

The third perturbation [represented by Fig. 1(C)] is specified by a step function
$[\theta(\rho-R_1)-\theta (\rho-R_2)]$ in front of Eqs. $(2.5) - (2.8)$. Then the black-box
cleavage operator $\tilde{V}_3(R_i,\rho')\delta(\rho'-R_i)P_{nn'}$ [with $P_{nn'}=1 (0)$ for
$n,n'\le 2$ and $\ge 3$ (otherwise); $i=1$ (2) for $n,n'\le 2$ ($\ge 3$)] is defined such that

\begin{equation}
\tilde{V}_3(R_i, \rho')=\frac{1}{2}\,\frac{q_{_{0}}^2}{\beta_3^2}
\left[
\begin{array}{cccc}
 \epsilon_3R_1 \frac{\partial}{\partial \rho'} \  &  \  \frac{imk}{q_{_{0}}\rho'}R_1 \  & \ 0 \ & \ 0 \\
-\frac{imk}{q_{_{0}}\rho'}R_1 \  & \ \mu_3 R_1\frac{\partial}{\partial \rho'} \ & \ 0 \ & \ 0 \\
0 \ & \ 0 \ & \ -\epsilon_3 R_2\frac{\partial}{\partial \rho'} \  &  \  -\frac{imk}{q_{_{0}}\rho'}R_2 \\
0 \ & \ 0 \ & \ \frac{imk}{q_{_{0}}\rho'}R_2 \  & \ -\mu_3 R_2\frac{\partial}{\partial \rho'}
\end{array}
\right]
\end{equation}
The corresponding bulk Green function is written as

\begin{eqnarray}
\tilde{G}_3(\rho, \rho')=i\pi\, \frac{\beta_3^2}{q_{_{0}}^2}
\left[
\begin{array}{cc}
-\frac{1}{\epsilon_3} J_m(\beta_3\rho) H_m(\beta_3\rho') \  & \  0 \\
0 \ & \  -\frac{1}{\mu_3}J_m(\beta_3\rho) H_m(\beta_3\rho') \\
-\frac{1}{\epsilon_3} H_m(\beta_3\rho) J_m(\beta_3\rho')  \ & \  0 \\
0 \ & \  -\frac{1}{\mu_3}H_m(\beta_3\rho) J_m(\beta_3\rho')
\end{array}
\right.\nonumber \\
\nonumber \\
\left.
\begin{array}{cc}
 -\frac{1}{\epsilon_3} J_m(\beta_3\rho) H_m(\beta_3\rho') \ & \ 0 \\
0 \ & \ -\frac{1}{\mu_3}J_m(\beta_3\rho) H_m(\beta_3\rho') \\
-\frac{1}{\epsilon_3} H_m(\beta_3\rho) J_m(\beta_3\rho') \ & \ 0 \\
0 \ & \ -\frac{1}{\mu_3}H_m(\beta_3\rho) J_m(\beta_3\rho')
\end{array}
\right]
\end{eqnarray}
where the interface space $M$ will be referred to ($\rho=R_1$, $\rho'=R_1$), ($\rho=R_1$, $\rho'=R_2$),
($\rho=R_2$, $\rho'=R_1$) and ($\rho=R_2$, $\rho'=R_2$), respectively, in the first, second, third,
and fourth quadrants made up of $2\times 2$ submatrices starting clockwise from the top-left.
With this, we define the response operator

\begin{eqnarray}
\tilde{A}_3(M,M)
=&&\tilde{V}_3(M)\,\tilde{G}_3(M,M)\nonumber\\
\nonumber\\
=&&\frac{i\pi}{2}\left[
\begin{array}{cc}
-\beta_3R_1J'_m(\beta_3R_1)H_m(\beta_3R_1) \ & \ -\frac{imk}{q_{_{0}}\mu_3}J_m(\beta_3R_1)H_m(\beta_3R_1)\\
\frac{imk}{q_{_{0}}\epsilon_3}J_m(\beta_3R_1)H_m(\beta_3R_1) \ & \ -\beta_3R_1J'_m(\beta_3R_1)H_m(\beta_3R_1)\\
\beta_3R_2H'_m(\beta_3R_2)J_m(\beta_3R_1) \ & \
\frac{imk}{q_{_{0}}\mu_3}H_m(\beta_3R_2)J_m(\beta_3R_1)\\
-\frac{imk}{q_{_{0}}\epsilon_3}H_m(\beta_3R_2)J_m(\beta_3R_1) \ & \ \beta_3R_2H'_m(\beta_3R_2)J_m(\beta_3R_1)
\end{array}
\right.\nonumber\\
\nonumber\\
&&\left.
\begin{array}{cc}
-\beta_3R_1J'_m(\beta_3R_1)H_m(\beta_3R_2) \ & \
-\frac{imk}{q_{_{0}}\mu_3}J_m(\beta_3R_1)H_m(\beta_3R_2)\\
\frac{imk}{q_{_{0}}\epsilon_3}J_m(\beta_3R_1)H_m(\beta_3R_2) \ & \ -\beta_3R_1J'_m(\beta_3R_1)H_m(\beta_3R_2)\\
\beta_3R_2H'_m(\beta_3R_2)J_m(\beta_3R_2) \ & \
\frac{imk}{q_{_{0}}\mu_3}H_m(\beta_3R_2)J_m(\beta_3R_2)\\
-\frac{imk}{q_{_{0}}\epsilon_3}H_m(\beta_3R_2)J_m(\beta_3R_2) \ & \ \beta_3R_2H'_m(\beta_3R_2)J_m(\beta_3R_2)
\end{array}
\right]
\end{eqnarray}
Now we define an operator

\begin{eqnarray}
\tilde{\Delta}_3(M,M)=&&
\tilde{I}\,+\,\tilde{A}_3(M,M) \nonumber \\
\nonumber\\
=&&\frac{i\pi}{2}\left[
\begin{array}{cc}
-\beta_3R_1H'_m(\beta_3R_1)J_m(\beta_3R_1) \ & \
-\frac{imk}{q_{_{0}}\mu_3}J_m(\beta_3R_1)H_m(\beta_3R_1)\\
\frac{imk}{q_{_{0}}\epsilon_3}J_m(\beta_3R_1)H_m(\beta_3R_1) \ & \ -\beta_3R_1H'_m(\beta_3R_1)J_m(\beta_3R_1)\\
\beta_3R_2H'_m(\beta_3R_2)J_m(\beta_3R_1) \ & \
\frac{imk}{q_{_{0}}\mu_3}H_m(\beta_3R_2)J_m(\beta_3R_1)\\
-\frac{imk}{q_{_{0}}\epsilon_3}H_m(\beta_3R_2)J_m(\beta_3R_1) \ & \ \beta_3R_2H'_m(\beta_3R_2)J_m(\beta_3R_1)
\end{array}
\right.\nonumber\\
\nonumber\\
&&\left.
\begin{array}{cc}
-\beta_3R_1J'_m(\beta_3R_1)H_m(\beta_3R_2) \ & \
-\frac{imk}{q_{_{0}}\mu_3}J_m(\beta_3R_1)H_m(\beta_3R_2)\\
\frac{imk}{q_{_{0}}\epsilon_3}J_m(\beta_3R_1)H_m(\beta_3R_2) \ & \ -\beta_3R_1J'_m(\beta_3R_1)H_m(\beta_3R_2)\\
\beta_3R_2J'_m(\beta_3R_2)H_m(\beta_3R_2) \ & \
\frac{imk}{q_{_{0}}\mu_3}H_m(\beta_3R_2)J_m(\beta_3R_2)\\
-\frac{imk}{q_{_{0}}\epsilon_3}H_m(\beta_3R_2)J_m(\beta_3R_2) \ & \ \beta_3R_2J'_m(\beta_3R_2)H_m(\beta_3R_2)
\end{array}
\right]
\end{eqnarray}
Again, in writing the second equality in Eq. (3.19), we have made use of the identity in Eq. (3.5). Next, we calculate the inverse of the bulk Green function $\tilde{G}_3$ to write

\begin{equation}
\tilde{G}^{-1}_3(M,M)=\frac{q_{_{0}}^2}{i\pi\beta_3^2}\,\frac{1}{D}
\left[
\begin{array}{cccc}
-\epsilon_3\frac{J_m(\beta_3R_2)}{J_m(\beta_3R_1)}  \ & \ 0 \ & \ \epsilon_3 \ & \ 0 \\
0 \ & \ -\mu_3\frac{J_m(\beta_3R_2)}{J_m(\beta_3R_1)} \ & \ 0 \ & \ \mu_3 \\
\epsilon_3 \ & \ 0 \ & \ -\epsilon_3\frac{H_m(\beta_3R_1)}{H_m(\beta_3R_2)} \ & \ 0 \\
0 \ & \ \mu_3 \ & \ 0 \ & \ -\mu_3\frac{H_m(\beta_3R_1)}{H_m(\beta_3R_2)} \\
\end{array}
\right]
\end{equation}
where the symbol $D$ is defined as

\begin{equation}
D = H_m(\beta_3R_1)J_m(\beta_3R_2) - J_m(\beta_3R_1)H_m(\beta_3R_2)
\end{equation}
Finally, we calculate the inverse response function of a cylindrical shell bounded by two black boxes

\begin{equation}
\tilde{g}^{-1}_3(M,M)= \tilde{\Delta}_3(M,M)\,\tilde{G}^{-1}_3(M,M)
\end{equation}
to write

\begin{equation}
\tilde{g}^{-1}_3(M,M)=\frac{q_{_{0}}^2}{2\beta_3^2}
\left[
\begin{array}{cccc}
\beta_3R_1\epsilon_3\frac{Z_1}{D} \ & \ \frac{imk}{q_{_{0}}} \ & \ -\frac{2i\epsilon_3}{\pi D} \ & \ 0\\
-\frac{imk}{q_{_{0}}}\ & \ \beta_3R_1\mu_3\frac{Z_1}{D}\ & \ 0 \ & \ -\frac{2i\mu_3}{\pi D}\\
-\frac{2i\epsilon_3}{\pi D} \ & \ 0 \ & \ \beta_3R_2\epsilon_3\frac{Z_2}{D}\ & \ -\frac{imk}{q_{_{0}}}\\
0 \ & \ -\frac{2i\mu_3}{\pi D} \ & \ \frac{imk}{q_{_{0}}} \ & \ \beta_3R_2\mu_3\frac{Z_2}{D}
\end{array}
\right]
\end{equation}
where

\begin{eqnarray}
\begin{array}{ll}
Z_1 = H'_m(\beta_3R_1)J_m(\beta_3R_2)-J'_m(\beta_3R_1)H_m(\beta_3R_2) \\
Z_2 = H'_m(\beta_3R_2)J_m(\beta_3R_1)-J'_m(\beta_3R_2)H_m(\beta_3R_1)
\end{array}
\end{eqnarray}
Having calculated the inverse response functions for the three perturbations, it becomes an easy task to deduce the dispersion relations for the plasmonic wave propagation in the {\em real} physical systems. These are (i) a
metamaterial (dielectric) cylinder embedded in a dielectric (metamaterial) and (ii) a metamaterial (dielectric) shell surrounded by two unidentical dielectrics (metamaterials), for example. It is noteworthy that even though we specify
the two geometries as we do, the {\em dielectric} there can be a {\em different} metamaterial. It should also be made clear once and for all that by dielectric we mean a conventional, nondispersive,  within which the $\epsilon$ and $\mu$ are {\em constant} parameters.

\subsection{Metamaterial (Dielectric) Cylinder Embedded in Dielectric (Metamaterial)}

Merger of perturbations A and B results into a geometry of a metamaterial (dielectric) cylinder embedded in a dielectric (metamaterial). As such, one can write $\tilde{g}^{-1}= \tilde{g}_1^{-1}+ \tilde{g}_2^{-1}$, where $\tilde{g}^{-1}$ is the inverse response function of a single cylinder in a semi-infinite medium. That means that formally the determinant of the sum of inverse response functions in Eqs. (3.8) and (3.15), with $R_1=R=R_2$, equated to zero, i.e.,

\begin{equation}
\left \vert \tilde{g}^{-1}(M,M)\right \vert =
\left \vert \tilde{g}^{-1}_1(M,M)+\tilde{g}^{-1}_2(M,M)\right \vert =0
\end{equation}
should yield the dispersion relation for plasmonic waves with a mixed (TM and TE) character in a single cylindrical geometry. The result is

\begin{equation}
\left \vert
\begin{array}{cc}
-\left[\frac{\epsilon_1}{\beta_1}\frac{J'_m(\beta_1R)}{J_m(\beta_1R)}-
       \frac{\epsilon_2}{\beta_2}\frac{H'_m(\beta_2R)}{H_m(\beta_2R)}\right] \ & \
-\frac{imk}{Rq_{_{0}}}\left(\frac{1}{\beta_1^2}-\frac{1}{\beta_2^2}\right) \\
\frac{imk}{Rq_{_{0}}}\left(\frac{1}{\beta_1^2}-\frac{1}{\beta_2^2} \right) \ & \
-\left[\frac{\mu_1}{\beta_1}\frac{J'_m(\beta_1R)}{J_m(\beta_1R)}-
       \frac{\mu_2}{\beta_2}\frac{H'_m(\beta_2R)}{H_m(\beta_2R)}\right]
\end{array}
\right \vert  = 0
\end{equation}
or

\begin{eqnarray}
\left[\frac{\epsilon_1}{\beta_1}\frac{J'_m(\beta_1R)}{J_m(\beta_1R)}-
      \frac{\epsilon_2}{\beta_2}\frac{H'_m(\beta_2R)}{H_m(\beta_2R)}\right]
\left[\frac{\mu_1}{\beta_1}\frac{J'_m(\beta_1R)}{J_m(\beta_1R)}-
      \frac{\mu_2}{\beta_2}\frac{H'_m(\beta_2R)}{H_m(\beta_2R)}\right]\nonumber\\
\nonumber\\
=\left( \frac{m}{R} \right)^2 \frac{k^2}{q_{_{0}}^2}
\left(\frac{1}{\beta_1^2}-\frac{1}{\beta_2^2}\right)^2
\end{eqnarray}
This expression is exactly identical to Eq. (107) in Ref. [59], which was obtained through the use of traditional
boundary conditions decades ago by Stratton. It is very important to understand that the only physical situation
where TM and TE modes can be decoupled in the cylindrical geometry is when (the Bessel index) $m=0$ and/or (the propagation constant) $k=0$. Then the TM modes are characterized by the nonzero $E_z$, $E_r$, and $H_{\theta}$
and TE modes by the nonzero $H_z$, $H_r$, and $E_{\theta}$. As such, Eq. (27) allows the decoupling of the TM (represented by the first square bracket equated to zero) and TE (specified by the second square bracket equated
to zero) modes. Next, we concentrate on the TM modes for studying, for example, plasmons in a slender wire made
up of the cylinder in the limit $R \rightarrow 0$.

\subsubsection{Quantum Wire in the Electric Quantum Limit}

For $m=0$, the TM modes are characterized by the following dispersion law:

\begin{equation}
\frac{\epsilon_1}{\beta_1}\frac{J_1(\beta_1R)}{J_0(\beta_1R)}-
      \frac{\epsilon_2}{\beta_2}\frac{H_1(\beta_2R)}{H_0(\beta_2R)}=0\ \  ,
\end{equation}
since $\zeta'_0=-\zeta_1$, with $\zeta_m \equiv J_m, H_m$. It is not difficult to prove that Eq. (3.28) is exactly identical to Eq. (18) in Ref. [60], which was also derived through the use of boundary conditions, and represents the plasmon dispersion for the classical dielectric waveguide. It is a simple matter to understand that in order to be
able to deduce some expected results for the planar interface we need to employ the large-argument limit (i.e., $R \rightarrow \infty$). Imposing asymptotic expansions of the Bessel functions for large arguments (i.e., when
$H_1/H_0=i$ and $J_1/J_0=-i$), we obtain $\epsilon_1\beta_2+\epsilon_2\beta_1=0$, which is a well-known general dispersion law for the TM modes propagating on an interface between two unidentical media characterized by dielectric functions $\epsilon_1$ and $\epsilon_2$ (see, for example, Ref. 1); here $\beta_1$ and $\beta_2$ serve as the decay constants for the respective media and have to be purely imaginary for the plasmon-polaritons.

Intuitively, a macroscopic metamaterial cylinder with a vanishingly small radius should mimic a fashionable quantum
wire and hence one would expect such a semiclassical methodology as treated here to reproduce the corresponding {\it intrasubband} plasmon dispersion. Using the lowest-order expansions of the involved Bessel functions for the small arguments one can cast Eq. (3.28) in the form:

\begin{equation}
\epsilon_1\,+\,\frac{2\epsilon_2}{\alpha_2R} \frac{K_1(\alpha_2R)}{K_0(\alpha_2R)}=0
\end{equation}
where $\alpha_2=(k^2-q_{_{0}}^2\epsilon_2\mu_2)^{1/2}$ refers to the decay constant in the outer background medium. For outer medium as a dielectric, small radius ($R \rightarrow 0$), and long wavelength limit, the modified Bessel functions $K_0$ and $K_1$ are both positive. Therefore in order to obtain a bonafide solution of this equation, the dielectric function $\epsilon_1$ must be negative. This means that in the local approximation only the frequencies below the screened plasma frequency should make sense. In the limit of small radius, Eq. (3.28) can also be written as

\begin{equation}
\omega = \omega_0 \cdot \beta_2 R \cdot \left\vert\ln (\beta_2 R) \right\vert^{1/2}
\end{equation}
where $\omega_0=(2\pi n_{_{B}}e^2/m^*\epsilon_{2L})^{1/2}$, $\epsilon_{2L}$ the background dielectric constant in the outer semi-infinite medium, and $n_B$ the effective 3D carrier density. In its present form, Eq. (3.30) is an exact analogue of Eq. (2.13) in Ref. [61], but generalized to include the retardation effect.

\subsection{Metamaterial (Dielectric) Shell Bounded by Two Unidentical Dielectrics (Metamaterials)}

In this section, we are motivated to study a physical system made up of two coaxial cylinders where we can have the metamaterial shell bounded by two unidentical dielectrics or a dielectric shell bounded by two unidentical metamaterials, in general. We will study diverse situations of practical interest. Methodologically, such a geometry becomes realizable
by summing up the inverse response functions calculated in Eqs. (3.8), (3.15), and (3.23) in the interface space $M$. One can write $\tilde{g}^{-1}= \tilde{g}_1^{-1}+ \tilde{g}_2^{-1}+ \tilde{g}_3^{-1}$, where $\tilde{g}^{-1}$ is the
response function of the finite cylindrical shell surrounded by two, in general, unidentical media. The dispersion relation for plasmonic waves in such a resultant structure is derived by equating the determinant of the total inverse
response function to zero, i.e.,

\begin{equation}
\left \vert \tilde{g}^{-1}(M,M)\right \vert =
\left \vert \tilde{g}^{-1}_1(M,M)+\tilde{g}^{-1}_2(M,M)+\tilde{g}^{-1}_3(M,M)\right \vert =0
\end{equation}
After some straightforward mathematical steps, we simplify Eq. (3.31) to write it explicitly in the compact form as follows.

\begin{equation}
\left \vert
\begin{array}{cccc}
-R_1 \left(\frac{\epsilon_1}{\beta_1}A_1-\frac{\epsilon_3}{\beta_3}C_1\right) \ & \
-\frac{imk}{q_{_{0}}}\left(\frac{1}{\beta_1^2}-\frac{1}{\beta_3^2}\right) \ & \
-\frac{2i\epsilon_3}{\pi\beta_3^2 D} \ & \  0 \\
\frac{imk}{q_{_{0}}}\left(\frac{1}{\beta_1^2}-\frac{1}{\beta_3^2}\right) \ & \
-R_1 \left(\frac{\mu_1}{\beta_1}A_1-\frac{\mu_3}{\beta_3}C_1\right)  &  0  &
-\frac{2i\mu_3}{\pi\beta_3^2 D} \\
-\frac{2i\epsilon_3}{\pi\beta_3^2 D} \ & \ 0 \  & \
R_2 \left(\frac{\epsilon_2}{\beta_2}A_2+\frac{\epsilon_3}{\beta_3}C_2\right) \ & \
\frac{imk}{q_{_{0}}}\left(\frac{1}{\beta_2^2}-\frac{1}{\beta_3^2}\right) \\
0 \ & \ -\frac{2i\mu_3}{\pi\beta_3^2 D} \  & \
-\frac{imk}{q_{_{0}}}\left(\frac{1}{\beta_2^2}-\frac{1}{\beta_3^2}\right) \ & \
R_2 \left(\frac{\mu_2}{\beta_2}A_2+\frac{\mu_3}{\beta_3}C_2\right) \\
\end{array}
\right \vert = 0
\end{equation}
where the additional substitutions are defined as

\begin{eqnarray}
\begin{array}{lll}
A_1 &=& J'_m(\beta_1R_1)/J_m(\beta_1R_1)\\
A_2 &=& H'_m(\beta_2R_2)/H_m(\beta_2R_2)\\
C_1 &=& Z_1/D\\
C_2 &=& Z_2/D
\end{array}
\end{eqnarray}
We are now interested to check how Eq. (3.32) can reproduce some well established results. For this purpose, we consider the limit $R_1 \sim R_2 \sim R \rightarrow \infty$ but take $R_2-R_1=d$ as a finite quantity and fix $m=0$. Naturally then, we need to make use of the asymptotic limits of the Bessel functions $J_{\nu}(z)$ and $H_{\nu}(z)$. As such, we first simplify the substitutions involved to obtain: $A_1=-i$, $A_2=i$, $D=-(2i/\pi\alpha_3R)\sinh(\alpha_3d)$,
$Z_1=Z_2=(2/\pi\alpha_3R)\cosh(\alpha_3d)$, and $C_1=C_2=i\coth(\alpha_3d)$; here $\beta_3=i\alpha_3$
just as before. As a consequence, we simplify the general dispersion relation in Eq. (3.32) to write

\begin{equation}
\left \vert
\begin{array}{cccc}
R \left(\frac{\epsilon_1}{\alpha_1}+\frac{\epsilon_3}{\alpha_3}C\right) \ & \
\frac{imk}{q_{_{0}}}\left(\frac{1}{\alpha_1^2}-\frac{1}{\alpha_3^2}\right) \ & \
-R \frac{\epsilon_3}{\alpha_3}S \ & \  0 \\
-\frac{imk}{q_{_{0}}}\left(\frac{1}{\alpha_1^2}-\frac{1}{\alpha_3^2}\right) \ & \
R \left(\frac{\mu_1}{\alpha_1}+\frac{\mu_3}{\alpha_3}C\right)  &  0  &
-R \frac{\mu_3}{\alpha_3}S \\
-R \frac{\epsilon_3}{\alpha_3}S \ & \ 0 \  & \
R \left(\frac{\epsilon_2}{\alpha_2}+\frac{\epsilon_3}{\alpha_3}C\right) \ & \
-\frac{imk}{q_{_{0}}}\left(\frac{1}{\alpha_2^2}-\frac{1}{\alpha_3^2}\right) \\
0 \ & \ -R\frac{\mu_3}{\alpha_3}S \  & \
\frac{imk}{q_{_{0}}}\left(\frac{1}{\alpha_2^2}-\frac{1}{\alpha_3^2}\right) \ & \
R \left(\frac{\mu_2}{\alpha_2}+\frac{\mu_3}{\alpha_3}C\right) \\
\end{array}
\right \vert = 0
\end{equation}
where $C=\coth\theta$ and $S^{-1}=\sinh\theta$; with $\theta=\alpha_3d$. Now let us carefully impose the limit $R \rightarrow \infty$. Then it is a simple matter to prove that Eq. (3.34) reduces to the form

\begin{equation}
\left[\frac{\epsilon_1\epsilon_2}{\alpha_1\alpha_2}+
\left(\frac{\epsilon_1}{\alpha_1}+
\frac{\epsilon_2}{\alpha_2}\right)\frac{\epsilon_3}{\alpha_3}\coth\theta +
\left(\frac{\epsilon_3}{\alpha_3}\right)^2\right]\\
\left[\frac{\mu_1\mu_2}{\alpha_1\alpha_2}+
\left(\frac{\mu_1}{\alpha_1}+\frac{\mu_2}{\alpha_2}\right)\frac{\mu_3}{\alpha_3}\coth\theta +
\left(\frac{\mu_3}{\alpha_3}\right)^2\right]=0
\end{equation}
Either the first or the second factor is zero. It is not quite difficult to prove that the first (second) factor equated to zero yields the TM (TE) modes propagating in the {\it planar} film geometry. Again, this form of the dispersion relation for the TE modes (as represented by the second factor equated to zero) is not, to the best of our knowledge, yet known in the literature. We focus on the first factor to study the 2D plasmons in a quantum well in the limit
$d \rightarrow 0$.

\subsubsection{Quantum Well in the Electric Quantum Limit}

The trigonometrical factor $\coth\theta$ in the limit of $\theta \rightarrow 0$ can be expanded in the following approximate form:
\begin{equation}
\coth\theta \,\simeq\, \frac{1}{\theta} + \frac{\theta}{3} - \frac{\theta^3}{45}
+ \frac{2\theta^5}{945}- \cdot  \cdot  \cdot  \cdot  \cdot
\end{equation}
where $\theta=\alpha_3d$. We consider the situation where $\epsilon_1$ and $\epsilon_2$ refer to the dielectric media and $\epsilon_3$ to the metamaterial. A simple mathematical analysis (in the limit $d \rightarrow 0$ and hence $\epsilon_3 \rightarrow \infty$) also leads us to deduce that $\alpha_3^2d^2 \rightarrow 0$ and $\epsilon_3d \simeq 4\pi\chi $,
with $\chi =-n_se^2/m^*\omega^2$ referring to the zero-temperature polarizability function in the long wavelength limit, remains a finite quantity. As such, retaining only the first term in the approximation and equating the first factor,
in Eq. (3.35), to zero yields
\begin{equation}
4\pi\chi +\frac{\epsilon_1}{\alpha_1}+\frac{\epsilon_2}{\alpha_2}=0
\end{equation}
This is now a well-known result that represents the plasma modes of a single 2DEG layer sandwiched between two dielectrics (see, for example, Ref. [1]). Furthermore, considering the bounding media to be identical (i.e., $\epsilon_1=\epsilon_2=\epsilon$ and hence $\alpha_1=\alpha_2=\alpha$) and imposing the non-retardation limit
(i.e., $q_{_{0}}=0$ and hence $\alpha=k$) leaves us with
\begin{equation}
\omega^2\,=\,\left(2\pi n_s e^2/m^* \epsilon\right)k
\end{equation}
where $n_s$ stands for the areal carrier density. This is a standard result for the intrasubband plasmon dispersion in
a quantum well, with the plasma frequency $\omega_p \propto \sqrt{k} $.

\subsection{Local and Total Density of States}

The density of states (DOS) is of fundamental importance to the understanding of many physical phenomena in condensed matter physics. Interpretation of quite a number of experimental excitation spectra in a wide variety of systems subjected to different physical conditions requires a detailed knowledge of the DOS. The classic textbooks and monographs reveal that the standard algorithm of determining the density of states is founded on the Green-function approach. Our purpose here is to calculate the local and total DOS in order to substantiate the computed plasmonic waves in the cylindrical geometries at hand. Unless some numeric hurdle comes in the way, this is logical to expect that the peaks in the DOS should coincide with the zeros of the inverse response function, which determine the plasmonic modes for a given propagation vector, of a system.

\subsubsection{Local Density of States}

The formal expression for the local density of states (LDOS) in the framework of interface response theory [53] is generally quite fussy and as the name suggests requires some subtle details of the local physical conditions. These
are, for example, the basic definitions of the bulk Green functions, the spatial positions around the interface, the nature of the associated EM fields involved, ... etc. In the present context, the simplest definition of the LDOS at
the expense of a few negligible concerns but that which still contains the important physics involved is given by
\begin{equation}
\mathcal{N}_{_{L}}(\omega) = -2\,\frac{\omega}{\pi}\,\mathrm{Im}\,
                                   \{\mathrm{trace}\,[\tilde{g}(M,M)]\}\\
\end{equation}
where $\tilde{g}$ refers to the response function whose inverse was determined in the preceding subsections for diverse situations. The important thing is to understand which system this response function $\tilde{g}$ refers to in different physical situations. We consider two such cases of our interest: a single-interface system (see Sec. III.D) and a double-interface system (see Sec. III.E). For a single-interface system, $\tilde{g}$ is simply the inverse of the sum of $\tilde{g}^{-1}_1$ and  $\tilde{g}^{-1}_2$ (see Sec. III.D). In the case of a two-interface system, we need to study the LDOS at the two interfaces $R_1$ and $R_2$ independently. For the interface $R_1$ ($R_2$) the $\tilde{g}$ in Eq. (3.39) is the $2 \times 2$ submatrix in the first (fourth) quadrant of the inverse of the sum of three inverse response functions (see Sec. III.E).

\subsubsection{Total Density of States}

For the z-components of the electromagnetic fields considered here, the analytical expression for the variation of the total density of states (TDOS) within the interface response theory [53] is given by
\begin{equation}
\mathcal{N}_{_{T}}(\omega) =\,-\, \frac{1}{\pi}\,\frac{d}{d\omega}\left( \mathrm{Arg}\,\, \mathrm{det}
       \left[ \frac{\tilde{g}_{_{i}}(M,M)}{\tilde{g}_{_{f}}(M,M)} \right]\right)\\
\end{equation}
By the variation of TDOS we mean the difference between the TDOS of the final (physical) system and an initial system. Here $\tilde{g}_{_{i}}$ ($\tilde{g}_{_{f}}$) stands for the response function of the initial (final) system in question. For the single-interface system, $\tilde{g}_i$ is a product of $\tilde{g}_1$ and $\tilde{g}_2$; and  $\tilde{g}_f$ is the inverse of the sum of $\tilde{g}^{-1}_1$ and $\tilde{g}^{-1}_2$. In the case of a two-interface system, $\tilde{g}_i=\tilde{g}_{1f}\cdot\tilde{g}_{2f}$, where $\tilde{g}_{1f}$ is the inverse of the sum of a $4 \times 4$ matrix comprised of the $\tilde{g}^{-1}_1$ and $\tilde{g}^{-2}_2$, and $\tilde{g}_{2f}$ is the inverse of $\tilde{g}^{-1}_3$ that corresponds to perturbation 3 for the shell alone; and $\tilde{g}_f$ is the inverse of the sum of $\tilde{g}^{-1}_1$, $\tilde{g}^{-1}_2$, and $\tilde{g}^{-1}_3$. It should be pointed out that both local and total DOS are computed for every value of integer $m$.

It is also worth mentioning that in the course of studying the total DOS we have the finite (or bounded) parts of the system automatically incorporated. Therefore, we are bound to find some discrete modes in the TDOS, which usually
appear as the negative peaks in the DOS$-\omega$ space and do not bear any physical significance if one is only interested in studying the confined or extended plasmon-polaritons. Moreover, if we are interested to understand all
the existing peaks in the TDOS, we need to explore, for example, each of the three perturbations involved individually. We have found that while the negative peaks in the individual perturbations survive in the TDOS, all the positive peaks are seen to disappear. This remains unfailingly true for all the cases we have investigated both for single- and double-interface systems. In a conventional system (made up of RHM), all the peaks in the LDOS are always positive. But in the present case where we have a system fabricated by interlacing the LHM with the RHM, the situation may take a different turn. More specific comments will be made later (see Sec. IV). It seems worthwhile to define the LHM as a special class of metamaterials where Re[$\epsilon$]$ < 0$ as well as Re[$\mu$]$ < 0$.

\subsection{Simple diagnoses of interface modes}

Here we make some simple analytical diagnoses of the confined modes propagating on an interface between a metamaterial and a dielectric. We intend to do so in order, for example, to understand the asymptotic limits, the slopes, the valid regions of their existence, and the physical conditions that allow or disallow the simultaneous existence of the TM and TE confined modes. In doing so we will distinguish between the non-dispersive and dispersive metamaterial components of the resultant structure.

\subsubsection{Non-dispersive metamaterial components}

\noindent {\it On the simultaneous existence of TM and TE modes}\\

By non-dispersive metamaterial we mean the LHM where $\epsilon$ and $\mu$ are both (negative) constants. Here we wish to address the following question: Can the TM and TE confined modes exist simultaneously? In order to answer this question, we recall Eq. (3.27) which in the limit of $R \rightarrow \infty$ splits into TM and TE modes. Let us first write explicitly
the result for the TM modes:
\begin{equation}
\frac{\epsilon_1}{\beta_1} + \frac{\epsilon_2}{\beta_2}=0
\end{equation}
This equation can be easily cast in the following form.
\begin{equation}
k^2=q_{_{0}}^2\,\, \frac{\frac{\mu_1}{\epsilon_1}-\frac{\mu_2}{\epsilon_2}}
                    {\frac{1}{\epsilon_1^2}-\frac{1}{\epsilon_2^2}}
\end{equation}
Remember for the true interface modes to exist [see, for instance, Ref. 1] (the propagation vector) $k$  has to be a real and positive quantity. Let us now analyze the case for the existence of TM modes. That means that both numerator and denominator have to have the same sign. This implies that either
\begin{equation}
\frac{\mu_1}{\epsilon_1} > \frac{\mu_2}{\epsilon_2} \ \ \ \ {\rm and} \ \ \ \
\frac{1}{\epsilon_1^2} > \frac{1}{\epsilon_2^2}
\end{equation}
or
\begin{equation}
\frac{\mu_1}{\epsilon_1} < \frac{\mu_2}{\epsilon_2} \ \ \ \ {\rm and} \ \ \ \
\frac{1}{\epsilon_1^2} < \frac{1}{\epsilon_2^2}
\end{equation}
Similarly, let us write the result for the TE modes:
\begin{equation}
\frac{\mu_1}{\beta_1} + \frac{\mu_2}{\beta_2}=0
\end{equation}
This equation can be written in the form
\begin{equation}
k^2=q_{_{0}}^2\,\, \frac{\frac{\epsilon_1}{\mu_1}-\frac{\epsilon_2}{\mu_2}}
                    {\frac{1}{\mu_1^2}-\frac{1}{\mu_2^2}}
\end{equation}
Again, for the true modes to exist the numerator and denominator have to have the same sign. This means that either
\begin{equation}
\frac{\epsilon_1}{\mu_1} > \frac{\epsilon_2}{\mu_2} \ \ \ \ {\rm and} \ \ \ \
\frac{1}{\mu_1^2} > \frac{1}{\mu_2^2}
\end{equation}
or
\begin{equation}
\frac{\epsilon_1}{\mu_1} < \frac{\epsilon_2}{\mu_2} \ \ \ \ {\rm and} \ \ \ \
\frac{1}{\mu_1^2} < \frac{1}{\mu_2^2}
\end{equation}
It is a simple matter to verify that each of Eqs. (3.43) and (3.44) remains always inconsistent with each of Eqs. (3.47) and (3.48). That is to say that they cannot be satisfied simultaneously. This leads us to infer that the TM and TE confined modes cannot exist simultaneously in the case that a non-dispersive metamaterial forms a component of a composite structure. While talking of the non-dispersive metamaterials sounds to be a fiction, it's a fact that many authors have explored some interesting physical aspects by treating them as nondispersive media describable by the
{\em (negative) constant} $\epsilon$ and $\mu$ [see, e.g., refs.$5-12$].\\

\noindent {\it Understanding the slopes}\\

There are different ways to make sure that the numerical results one has obtained are in reasonably good correspondence with the analytical results. One of them is to examine, for example, the slopes of the coupled and/or decoupled modes. As we now know, even the coupled modes (with $m \neq 0$) in the asymptotic limit ($k \rightarrow \infty$ and/or $R \rightarrow \infty$) become asymptotic to the decoupled modes. Then it becomes interesting to sometimes examine their slopes. In this situation, one can rewrite Eqs. (3.42) and (3.46) as
\begin{equation}
\frac{\omega^2}{c^2k^2}=\,
\frac{\frac{1}{\epsilon_1^2}-\frac{1}{\epsilon_2^2}}
{\frac{\mu_1}{\epsilon_1}-\frac{\mu_2}{\epsilon_2}}
\end{equation}
for the TM modes and
\begin{equation}
\frac{\omega^2}{c^2k^2}=\,
\frac{\frac{1}{\mu_1^2}-\frac{1}{\mu_2^2}}
{\frac{\epsilon_1}{\mu_1}-\frac{\epsilon_2}{\mu_2}}
\end{equation}
for the TE modes. Equations (3.49) and (3.50) define the slopes of the TM and TE modes in the asymptotic limit. These equations prove to be cautionary particularly in the situation where one may choose some special set of parameters such as $\epsilon_2=\pm\,\epsilon_1$ and/or $\mu_2=\pm\, \mu_1$. In this case one may be surprised to notice that not only these equations dictate the slopes to be zero, but also that one may miss the confined modes altogether.

\subsubsection{Dispersive metamaterial components}

\noindent {\it Asymptotic limits}\\

By dispersive metamaterial we mean the LHM where both $\epsilon$ and $\mu$ are the functions of frequency. We choose them such as follows:
\begin{equation}
\epsilon(\omega) = 1 - \frac{\omega_p^2}{\omega^2}
\end{equation}
where $\omega_p$ is the plasma frequency (usually in the GHz range) and
\begin{equation}
\mu(\omega) = 1 - \frac{F\omega^2}{\omega^2-\omega_0^2}
\end{equation}
where $F$ is a constant factor and $\omega_0$ is the resonance frequency. The simplest check on the numerical results
is to examine the asymptotic limits (i.e., $k\rightarrow \infty$) the confined modes are approaching to. In the case that medium $1$ is a nondispersive RHM and medium $2$ is a dispersive LHM, Eq. (3.41) yields
\begin{equation}
W=\frac{\omega}{\omega_p}=\frac{1}{\sqrt{\epsilon_1 +1}}
\end{equation}
and Eq. (45) yields
\begin{equation}
W=\frac{\omega}{\omega_p}=\frac{W_0}{\sqrt{1-\frac{F}{\mu_1+1}}}
\end{equation}
where $W_0=\omega_0/\omega_p$. For a specific set of parameters, Eqs. (3.53) and (3.54), respectively,  determine the asymptotic limits of the TM and TE modes.\\

\noindent{\it Valid regions for the confined modes}\\

It is a simple matter to analyze that $\epsilon (\omega) < 0$ for $W < 1$ and $\mu (\omega)<0$ for
$W<W_c$, where $W_c=W_0/\sqrt{1-F}$. Region $W<W_0$: $\epsilon(\omega)<0$ and $\mu(\omega)>0$ $\Rightarrow$ $\beta_{LHM}$ is pure imaginary $\Rightarrow$ $\alpha_{LHM}=-i\beta_{LHM}$ is pure real
and positive and hence a valid region for confined modes. Region $W_0 \le W\le W_c$:
$\epsilon(\omega)<0$ and $\mu(\omega)<0$ $\Rightarrow$ $\beta_{LHM}$ is real $\Rightarrow$
$\alpha_{LHM}$ is pure imaginary and hence no confined modes are allowed. Region $W_C \le W\le 1$:
$\epsilon(\omega)<0$ and $\mu(\omega)>0$ $\Rightarrow$ $\beta_{LHM}$ is pure imaginary $\Rightarrow$
$\alpha_{LHM}=-i\beta_{LHM}$ is real and positive and hence a valid region for confined modes.

It is worth mentioning that the resonance frequency $\omega_0$ generally allows a large number of unphysical solutions piled up in the immediate vicinity of $\omega_0$ in the $\omega-k$ space, which usually need to be eliminated in order to avoid any confusion. Something similar is seen to happen in the computation of the density of states.

\section{ILLUSTRATIVE EXAMPLES}

As we have seen in the preceding section, our final results for the dispersion characteristics are Eqs. (3.26) and (3.32), respectively, for the single cylindrical cable embedded in some different material background and the coaxial cylindrical system made up of a finite shell bounded by the closed (innermost) cable and the semi-infinite medium. Note that both of these equations are, in general, the complex transcendental functions. Therefore, in principle, we need to search the zeros of such complex functions.
We had to strike a compromise among a few choices. We decided to ask the machine to produce those zeros where the imaginary (real) part of the function changes the sign, irrespective of whether or not the real (imaginary) part is zero in the radiative (nonradiative) region.  We believe this has resulted into a reliable scheme for studying the dispersion characteristics of plasmonic waves in the present systems. This is because all the plasmonic waves (confined or extended) are found to have excellent correspondence with the peaks in the local and/or total density of states. We purposely consider only the cases with dispersive metamaterials interlaced with conventional dielectrics (usually vacuum with $\epsilon =1=\mu$). The detailed studies of nondispersive metamaterials is deferred to a future publication. So the only parameters involved in the treatment are $F$ and $\omega_0$ and we choose them such that $F=0.56$, $\omega_p/2\pi=10$ GHz and $\omega_0/2\pi=4$ GHz; the latter yields $\omega_0/\omega_p=0.4$ [see Ref. 7]. We will later assign an additional numeral as a suffix to the background dielectric constants corresponding to the region in the geometry concerned. Other parameters such as the ratio of the radii of the cylinders $R_2/R_1$, the normalized plasma frequency $\omega_pR/c$, and the azimuthal index of the Bessel functions $m$ will be given at the appropriate places during the discussion. We will present our results in terms of the dimensionless propagation vector $\zeta=ck/\omega_{p}$ and frequency $\xi=\omega/\omega_{p}$, where $\omega_p$ stands for the screened plasma frequency. Both local and total DOS will be shown in arbitrary units throughout.

\subsection{Single-interface systems}

\begin{figure}[htbp]
\includegraphics*[width=9cm,height=10cm]{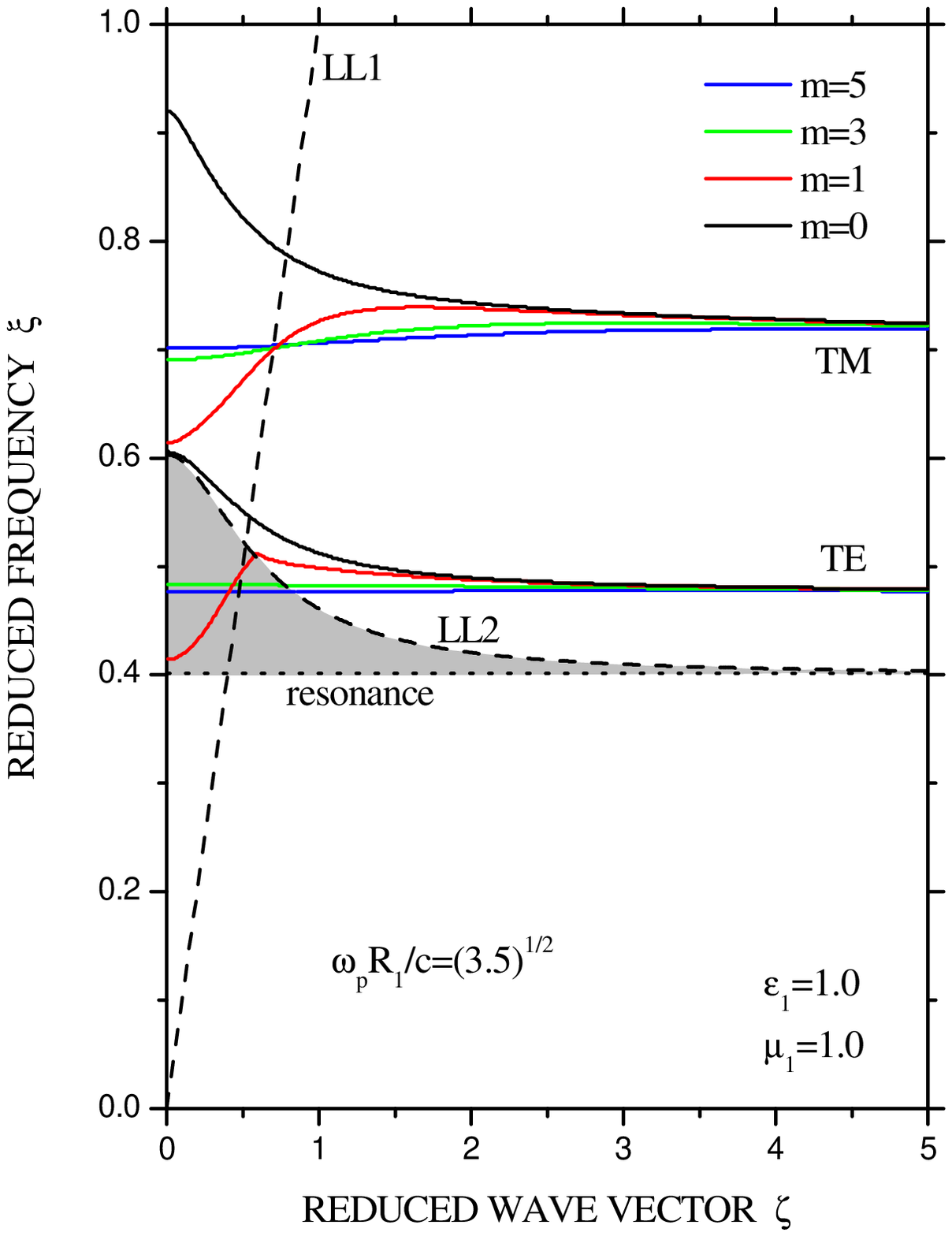}
\caption{(Color online) Plasmon dispersion for a dielectric (vacuum) cable embedded in a metamaterial background for different values of index $m=0$, 1, 3, 5. The dimensionless plasma frequency used in the computation is specified by $\omega_pR_1/c=\sqrt{3.5}$. Dashed line and curve marked as LL1 and LL2 refer, respectively, to the light lines in the vacuum and the metamaterial. The horizontal dotted line stands for the characteristic resonance frequency ($\omega_0$) 
in he metamaterial. The shaded area represents the region within which both $\epsilon(\omega)$ and $\mu(\omega)$ are negative and prohibits  the existence of the confined modes.}
\label{fig2}
\end{figure}

Figure 2 illustrates the plasmonic wave dispersion for the dielectric (vacuum) cable embedded in a metamaterial background
for $m=0$, 1, 3, and 5. The plots are rendered in terms of the dimensionless frequency $\xi$ and the dimensionless propagation vector $\zeta$.  The important parameter involved is the dimensionless plasma frequency specified by $\omega_pR_1/c=\sqrt{3.5}$. The dashed line and the curve marked as LL1 and LL2 refer, respectively, to the light
lines in the vacuum and the metamaterial. The horizontal dotted line stands for the characteristic resonance frequency ($\omega_0=0.4 \omega_p$) in the metamaterial. The shaded area represents the region where $\epsilon(\omega)< 0$ and $\mu(\omega)< 0$ and disallows the existence of the confined modes (see Sec. III.G). It is important to notice that the confined modes can be distinguished as TM or TE only when the Bessel function index $m=0$. However, we designate one group of modes as TM and another as TE, even for $m \ne 0$, simply on the basis of the asymptotic limits attained by them. Notice that both asymptotic limits are correctly dictated by Eqs. (3.53) and (3.54). What is noteworthy here is that the system supports the simultaneous existence of TM and TE modes. It is also interesting to notice that the resonance frequency $\omega_0$ is not seen to play any of its characteristic role in the spectrum (see in what follows).


\begin{figure}[htbp]
\includegraphics*[width=9cm,height=10cm]{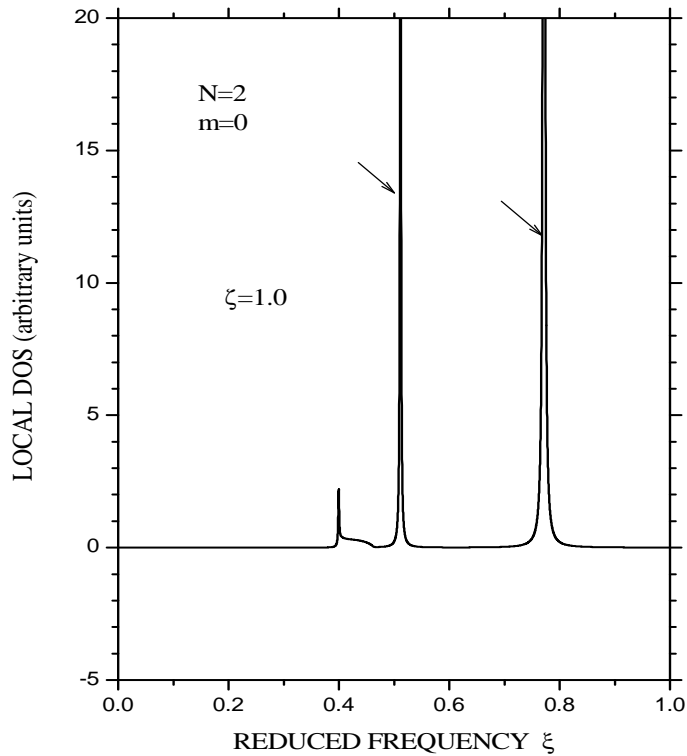}
\caption{Local density of states for the system discussed in Fig. 2 and for $m=0$ and $\zeta=1.0$. The rest of 
the parameters used are the same as in Fig. 2. The arrows in the panel indicate the peaks at $\xi=0.5119$ and
$\xi=0.7718$.}
\label{fig3}
\end{figure}

Figure 3 shows the local density of states for the system discussed in Fig. 2 for $m=0$ and $\zeta=1.0$. This value of the propagation vector lies in the non-radiative region which allows the pure confined modes. The rest of the parameters are the same as in Fig. 2. Both sharp peaks occurring at $\xi=0.5119$ and $\xi=0.7718$ reproduce exactly the respective TE and TM modes existing at $\zeta=1.0$ in Fig. 2. The short peak and the related noisy part in the immediate vicinity of the resonance frequency $\omega_0$ has no physical significance and will show its signature almost everywhere in the computation of local as well as total density of states. It has been found that similar calculation of LDOS at any value of the propagation vector correctly reproduces both modes in spectrum discussed in Fig. 2.


\begin{figure}[htbp]
\includegraphics*[width=9cm,height=10cm]{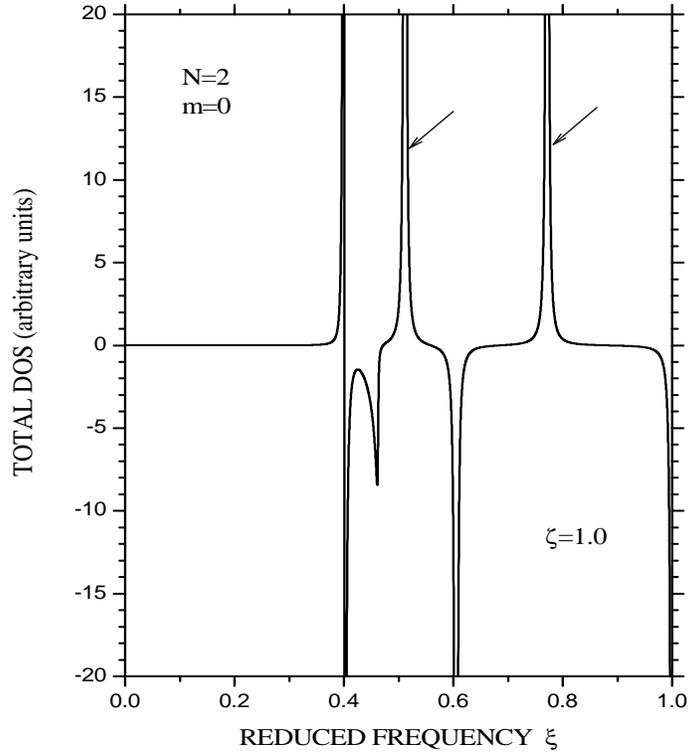}
\caption{Total density of states for the system discussed in Fig. 2 and for $m=0$ and $\zeta=1.0$. The rest of the parameters used are the same as in Fig. 2. Both negative peaks are characteristic of the
resonance frequency $\omega_0$ and other characteristic frequency $\omega_c$ in the system and bear no physical significance.}
\label{fig4}
\end{figure}

Figure 4 depicts the total density of states for the dielectric (vacuum) cable embedded in a dispersive metamaterial background discussed in Fig. 2 for $m=0$ and $\zeta=1.0$. For $m0$, the otherwise coupled modes are decoupled as TM and TE. The $\zeta=1.0$ indicates the positions of the TE and TM modes lying at $\xi=0.5119$ and $\xi=0.7718$ (see Fig. 2). Both of these positions of the respective modes are exactly reproduced by the peaks marked by arrows in the total density of states here. A kind of resonant behavior at $\xi\simeq 0.4$ and the negative peak at $\xi\simeq 0.6$ are characteristic of the critical frequencies $\omega_0$ and $\omega_c$ involved in the problem and we do not consider them of any importance and/or physical significance. Similar behavior at these frequencies will be seen in the later examples as well.


\begin{figure}[htbp]
\includegraphics*[width=9cm,height=10cm]{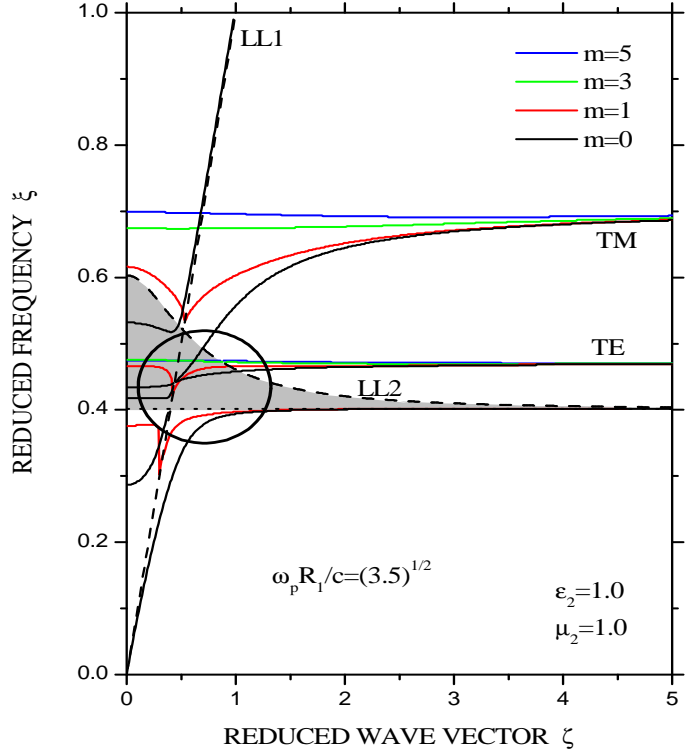}
\caption{(Color online) Plasmon dispersion for a metamaterial cable in a dielectric (vacuum) background for different values of index $m=0$, 1, 3, 5. The dimensionless plasma frequency used in the computation is specified by $\omega_pR_1/c=\sqrt{3.5}$. Dashed line and curve marked as LL1 and LL2 refer, respectively, to the light lines in the vacuum and the metamaterial. The horizontal dotted line stands for the characteristic resonance frequency ($\omega_0$) 
in the metamaterial. The shaded area represents the region within which both $\epsilon(\omega)$ and $\mu(\omega)$ are negative and disallows the existence of the confined modes.}
\label{fig5}
\end{figure}

Figure 5 illustrates the plasmonic wave dispersion for a metamaterial cable in a dielectric (vacuum) background for different values of index $m=0$, 1, 3, 5. The results are plotted in terms of the dimensionless frequency $\xi$ and the dimensionless propagation vector $\zeta$. The dimensionless plasma frequency used in the computation is specified by $\omega_pR_1/c=\sqrt{3.5}$. The dashed line and curve marked as LL1 and LL2 refer, respectively, to the light lines in the vacuum and the metamaterial. The horizontal dotted line stands for the characteristic resonance frequency ($\omega_0$) in the metamaterial. The shaded area represents the region where both $\epsilon(\omega)$ and $\mu(\omega)$ are negative and disallows the existence of confined modes. We designate the two groups of modes as TM and TE with the same notion as discussed in Fig. 2. One can see it clearly that the resonance frequency $\omega_0$ does play a crucial role in this case. For instance, the big hollow circle encloses the $m=0$ and the $m=1$ TM modes split due to the resonance frequency in the problem. We call this splitting occurring between respective TM modes since we can see it happening just by plotting the $m=0$ modes. If it were not for the resonance frequency the split $m=0$ mode would start from zero (just as here) and propagate smoothly to approach the asymptotic limit without any splitting and $m=1$ mode would emerge from a nonzero frequency somewhere in the radiative region. This is to stress that such resonance splitting takes place only for the TM modes and not for the TE modes. The latter always start from above the resonance frequency. It is noticeable that the split modes below $\omega_0$ later become asymptotic to $\omega_0$.  The TM modes' splitting behavior will become more transparent in the later examples on double-interface systems (see, for example, Figs. 8
and 11).


\begin{figure}[htbp]
\includegraphics*[width=9cm,height=10cm]{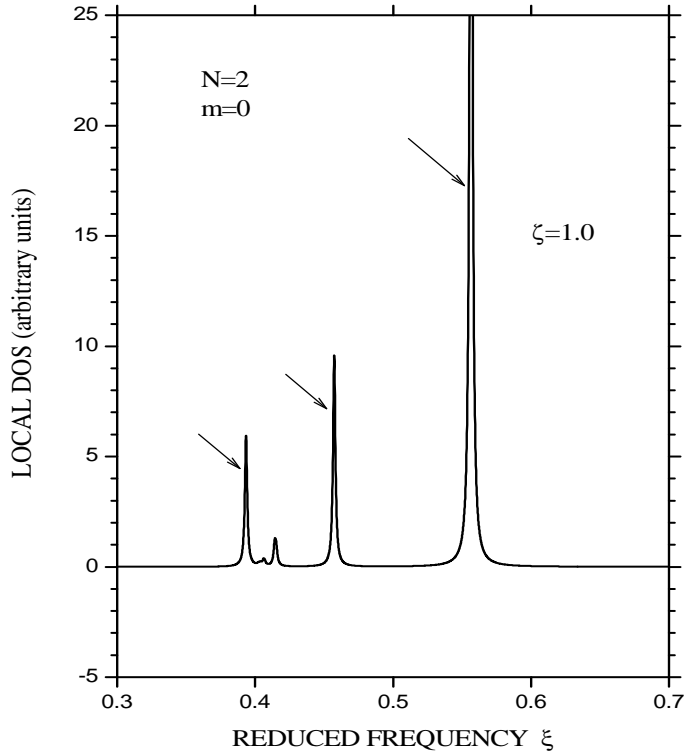}
\caption{Local density of states for the system discussed in Fig. 5 and for $m=0$ and $\zeta=1.0$. The 
arrows in the panel indicate the peaks at $\xi=0.3947$, $\xi=0.4581$, and  $\xi=0.5577$. The rest of
the parameters used are the same as in Fig. 5.}
\label{fig6}
\end{figure}

Figure 6 shows the local density of states for the system discussed in Fig. 5 for $m=0$ and $\zeta=1.0$. This value of the propagation vector lies in the non-radiative region which allows the pure confined modes. The rest of the parameters are the same as in Fig. 5. All the three sharp peaks lying at $\xi=0.3947$, $\xi=0.4581$, and  $\xi=0.5577$ reproduce exactly the respective TE and TM modes existing at $\zeta=1.0$ in Fig. 5. The short peak and the related noise in the immediate vicinity of the resonance frequency $\omega_0$ has no physical significance in the problem. It has been found that similar calculation of LDOS at any value of the propagation vector and for any given $m$ correctly reproduces
all the modes in spectrum discussed in Fig. 5. It should be pointed out that while the LDOS (and/or TDOS, for that matter) can and do reproduce lowest (TM) split mode at the higher propagation vector where this mode has already become asymptotic to and merged with $\omega_0$, sometimes it becomes extremely difficult to discern such a peak inside the band of noise existing in the immediate vicinity of $\omega_0$.


\begin{figure}[htbp]
\includegraphics*[width=9cm,height=10cm]{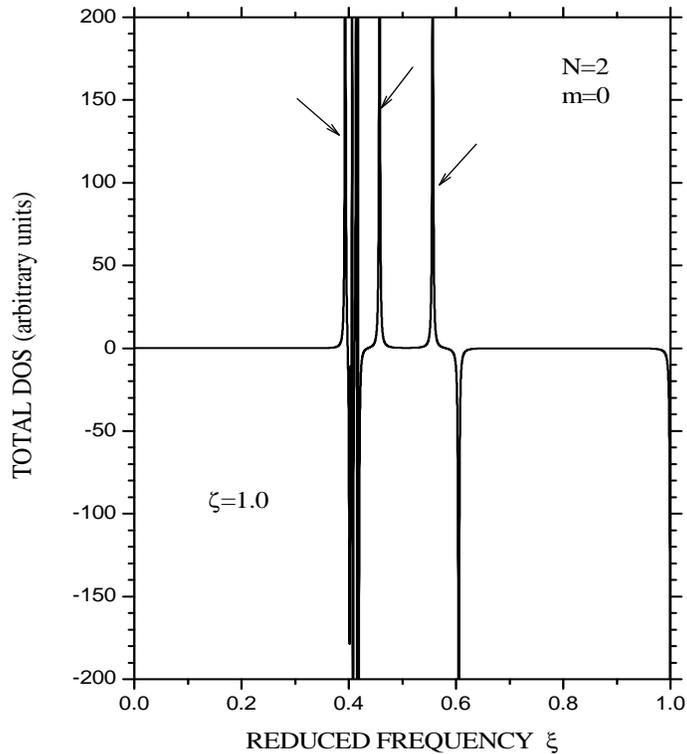}
\caption{Total density of states for the system discussed in Fig. 5 and for $m=0$ and $\zeta=1.0$. The 
arrows in the panel indicate the peaks at $\xi=0.3947$, $\xi=0.4581$, and  $\xi=0.5577$. The rest of
the parameters used are the same as in Fig. 5.}
\label{fig.7}
\end{figure}

Figure 7 depicts the total density of states for the metamaterial cable embedded in a dispersive dielectric (vacuum) background discussed in Fig. 5 for $m=0$ and $\zeta=1.0$. For $m0$, the otherwise coupled modes are decoupled as TM and TE. The $\zeta=1.0$ specifies the positions of the decoupled modes lying at $\xi=0.3947$, $\xi=0.4581$, and  $\xi=0.5577$. All of these positions of the respective modes are correctly reproduced by the peaks marked by the arrows in the total density of states here. The noisy band of states at $\omega_0$ and the negative peak at $\omega_c$ are a consequence of these critical frequencies involved in the problem but they carry no interesting information and bear no physical significance. Scanning the whole range of propagation vector reveals that the TDOS reproduces all the modes in Fig. 5 very accurately. The only exception to this is the radiative modes (towards the left of the light line) in Fig. 5, which do not show a good correspondence with the resonance peaks in the (local or total) DOS. This is not surprising, however, given the distinct ways of searching the zeros of the complex transcendental function in the radiative and non-radiative regions. We did not intend to pay much attention to the small radiative region simply because, as we all know, this region is of almost no practical interest.

\subsection{Double-interface systems}

\begin{figure}[htbp]
\includegraphics*[width=9cm,height=10cm]{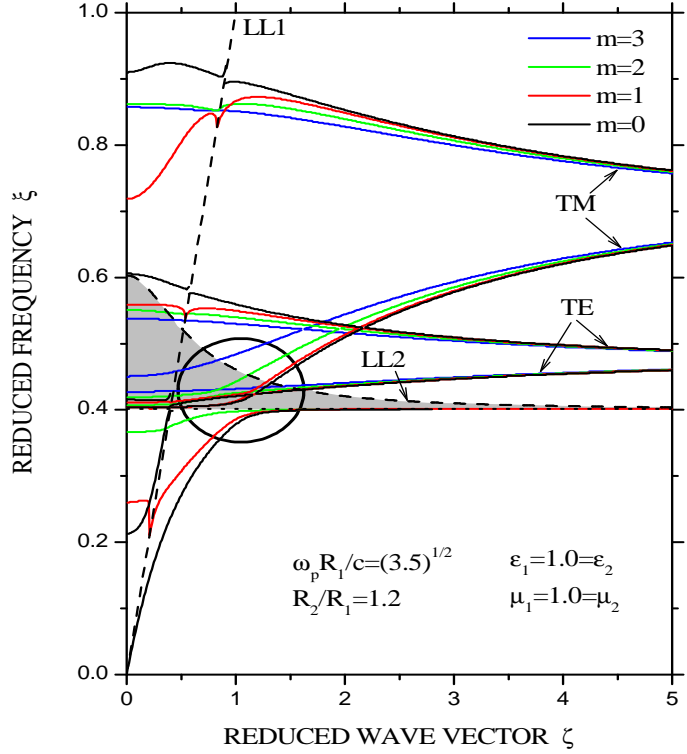}
\caption{(Color online) Plasmon dispersion for a metamaterial shell sandwiched between two identical dielectrics 
(vacuum) for different values of index $m=0$, 1, 2, 3. The dimensionless plasma frequency used here is specified by $\omega_pR_1/c=\sqrt{3.5}$ and the radii ratio $R_2/R_1=1.2$. Dashed line and curve marked as LL1 and LL2 refer, respectively, to the light lines in the vacuum and the metamaterial. The horizontal dotted line stands for the characteristic resonance frequency ($\omega_0$) in the metamaterial. The shaded area represents the region within which both $\epsilon(\omega)$ and $\mu(\omega)$ are negative and forbids the existence of the confined modes. The parameters used in the computation are as listed in the picture.}
\label{fig8}
\end{figure}

Figure 8 illustrates the plasmonic wave dispersion for a metamaterial shell sandwiched between two identical dielectrics (vacuum) for different values of index $m=0$, 1, 2, and 3. The plots are rendered in terms of the dimensionless frequency $\xi$ and the propagation vector $\zeta$. The other important parameters used in the computation are the normalized plasma frequency $\omega_pR_1/c=\sqrt{3.5}$ and the ratio $R_2/R_1=1.2$. The dashed line and the curve
marked as LL1 and LL2 refer, respectively, to the light lines in the bounding media (vacuum) and the metamaterial
shell. The horizontal dotted line refers to the characteristic resonance frequency ($\omega_0$) in the problem. The shaded area refers to the region where $\epsilon(\omega) < 0$ and $\mu(\omega) < 0$ and prohibits the existence of
the truly confined modes. Since there are two interfaces involved in the resultant structure we have a pair of modes
for each of the TM and TE modes in the system. The lower and upper group of modes together attain the same asymptotic limit characteristic of the TM or the TE modes at large wave vectors. The resonance frequency allows the splitting of the $m=0$, $m=1$, and $m=2$ TM modes at $\omega_0$. The full circle encloses and shows such a resonance splitting occurring between the respective modes in a very clear way at $\zeta \simeq 1.0$. The scheme of assigning the modes a
TM or a TE character is the same as discussed before (see discussion of Fig. 2). Eqs. (3.53) and (3.54) are seen to dictate the correct asymptotic limits attained both by TM and TE modes. Some abruptness (sharp or blunt) observed by a given mode at the light line is a general tendency usually seen when a mode crosses the junction between the two media.


\begin{figure}[htbp]
\includegraphics*[width=9cm,height=10cm]{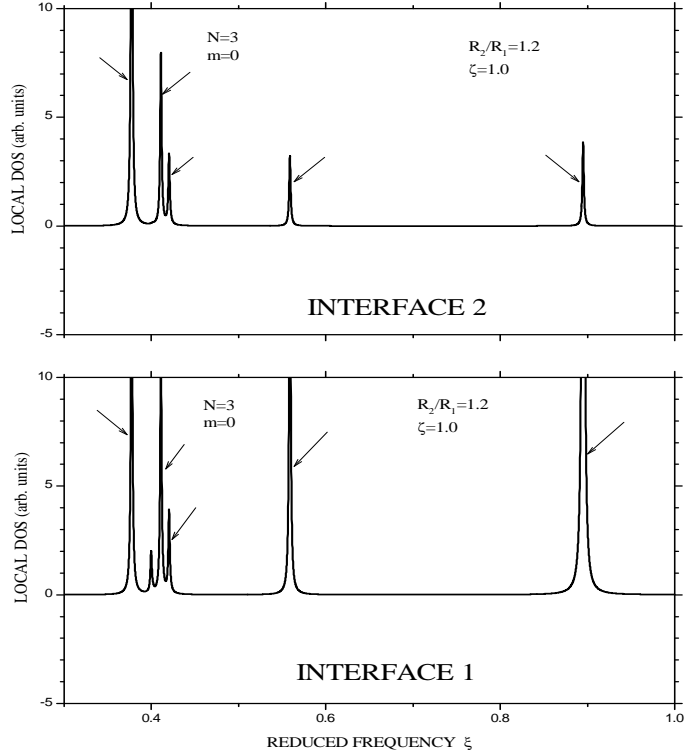}
\caption{Local density of states at the interface $R_1$ ($R_2$) in the lower (upper) panel for $m=0$ and
$\zeta=1.0$ for the system discussed in Fig. 8. We call attention to the DOS resonance peaks, indicated by the arrows, corresponding to the five modes in total at $\zeta=1.0$ in Fig. 8. The interface 1 (2) refers to the one specified by $R_1$ ($R_2$). The rest of the parameters used are the same as in Fig. 8.}
\label{fig9}
\end{figure}

Figure 9 shows the local density of states for the two-interface system discussed in Fig. 8 for $m=0$ and $\zeta=1.0$ for the interface 1 (2) in the lower (upper) panel. This value of $\zeta$ specifies five propagating modes in total in Fig. 8: the lowest split (TM) mode below $\omega_0$, lower (split) TM mode and lower TE mode within the shaded region, upper TE mode, and the uppermost TM mode lying, respectively, at $\xi=0.3774$, $\xi=0.4109$, $\xi=0.4202$, $\xi=0.5588$, and $\xi=0.8951$. In the lower panel, the five resonance peaks (indicated by arrows) observed in the local density of states stand exactly at these frequencies. This implies reasonably a very good correspondence between the (dispersion) spectrum and the LDOS at interface $R_1$. The (unmarked) second lowest peak (counting from the lowest frequency) stands at the resonance frequency $\omega_0$ and is not considered to be a bonafide peak in the LDOS. Coming to the upper panel, we again observe five well-defined resonance peaks lying exactly at the aforementioned frequencies. That means that both interfaces share all the five resonances in the LDOS, albeit with a difference of magnitude. This also implies that the two interfaces pose different preferences, and that makes sense here because of the asymmetric nature of the configuration. In other words, the two interfaces seem to be more sensitive to the geometry and less to the materials in the supporting media. That is to say that the situation is altogether different from a planar geometry with, for example, a thin metallic or semiconducting film symmetrically bounded by two identical dielectrics.


\begin{figure}[htbp]
\includegraphics*[width=9cm,height=10cm]{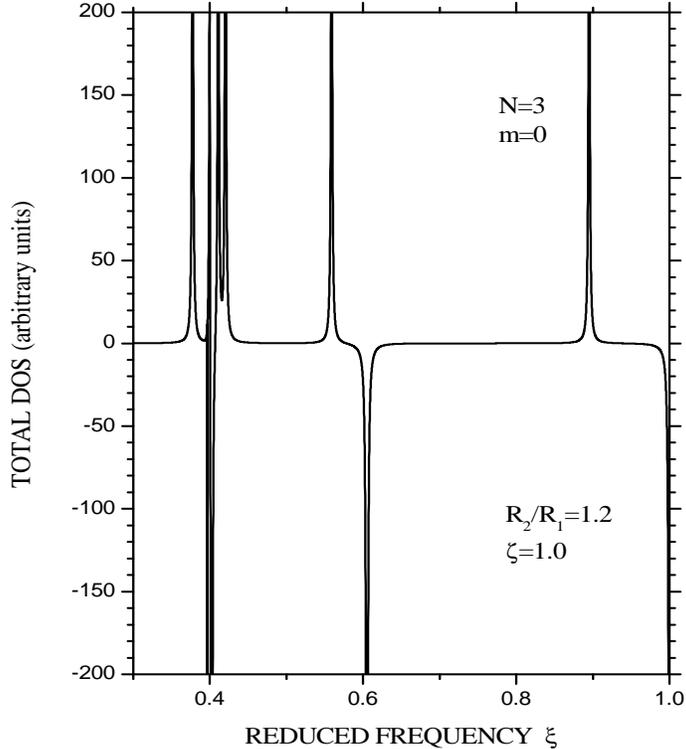}
\caption{Total density of states for $m=0$ and $\zeta=1.0$ for the system discussed in Fig. 8. We call attention to the DOS resonance peaks, indicated by the arrows, corresponding to the five modes in total at $\zeta=1.0$ in Fig. 8. The parameters used are the same as in Fig. 8.}
\label{fig10}
\end{figure}

Figure 10 depicts the total density of states for the two-interface system discussed in Fig. 8 for $m=0$ and $\zeta=1.0$. These values of $m$ and $\zeta$ define five propagating modes in total in Fig. 8, covering both TM and TE modes, and lying at $\xi=0.3774$, $\xi=0.4109$, $\xi=0.4202$, $\xi=0.5588$, and $\xi=0.8951$. The five resonance peaks (indicated by arrows) observed in the total density of states are seen to substantiate these frequencies very accurately. We have seen that the similar computation of TDOS at any other value of $\zeta$ reproduces all the corresponding modes in the spectrum very correctly. The only exception to this is the radiative region (toward the left of the light line) where the correspondence between the DOS (local or total) and the spectrum is no so good. This is again understandable in the view of the facts stated above (see the discussion of Fig. 7). A pile up of the states at and in the vicinity of $\omega_0$ and a negative peak at $\omega_c$ are clearly a consequence of the presence of such critical (resonance) frequencies in the problem and we do not consider them to be of any physical significance.


\begin{figure}[htbp]
\includegraphics*[width=9cm,height=10cm]{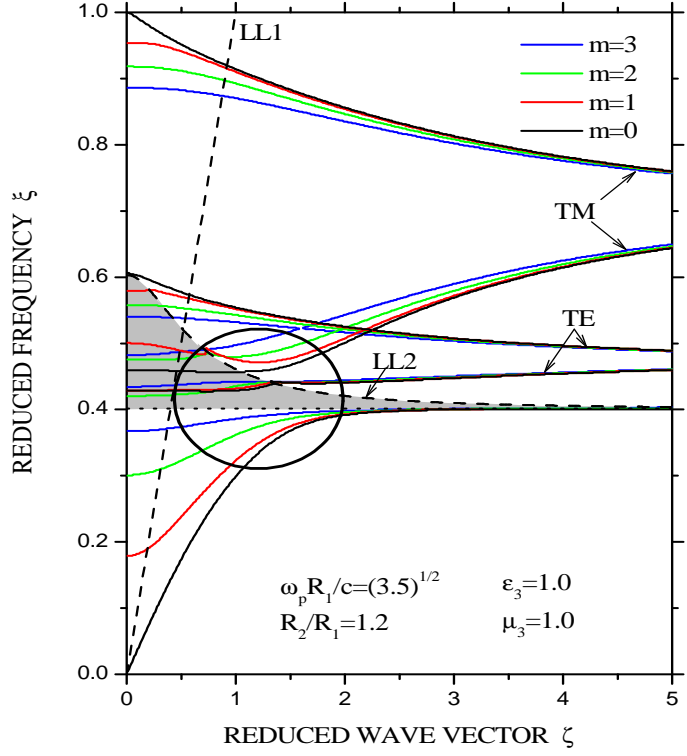}
\caption{(Color online) Plasmon dispersion for a dielectric (vacuum) shell sandwiched between two identical 
metamaterials for different values of index $m=0$, 1, 2, 3. The dimensionless plasma frequency used here is specified
by $\omega_pR_1/c=\sqrt{3.5}$ and the radii ratio $R_2/R_1=1.2$. Dashed line and curve marked as LL1 and LL2 refer, respectively, to the light lines in the vacuum and the metamaterial. The horizontal dotted line stands for the characteristic resonance frequency ($\omega_0$) in the metamaterial. The shaded area represents the region within
which both $\epsilon(\omega)$ and $\mu(\omega)$ are negative and proscribes the existence of the confined modes. The parameters used in the computation are as listed in the picture.}
\label{fig11}
\end{figure}

Figure 11 illustrates the plasmonic wave dispersion for a dielectric (vacuum) shell sandwiched between two identical metamaterials for different values of index $m=0$, 1, 2, and 3. The plots are rendered in terms of the dimensionless frequency $\xi$ and the propagation vector $\zeta$. The other important parameters used in the problem are the normalized plasma frequency $\omega_pR_1/c=\sqrt{3.5}$ and the ratio of the radii $R_2/R_1=1.2$. The dashed line and
the curve marked as LL1 and LL2 stand, respectively, for the light lines in the dielectric (vacuum) and the bounding metamaterials. The horizontal dotted line refers to the characteristic resonance frequency ($\omega_0$) in the problem. The shaded area refers to the region within which $\epsilon(\omega) < 0$ and $\mu(\omega) < 0$ and proscribes the existence of the truly confined modes. Again, since there are two interfaces in the system we obtain two pairs of modes: one for the TM and the other for the TE modes. Their asymptotic limits are governed by Eqs. (3.53) and (3.54). The presence of the resonance frequency $\omega_0$ gives rise to the resonance splitting of all the lower group of the pair of TM modes for different values of $m$. Although the lower group of the pair of TE modes cross in between the split TM modes (inside the shaded region), the resonance splitting is clearly pronounced between the corresponding TM modes. This is shown by the big hollow circle encompassing all the respective split TM modes in the region. Again, the scheme of assigning the modes a TM or a TE character is the same as discussed before. Notice that the abruptness observed by the modes while crossing the light line is relatively smoother than that seen in the other cases (cf. Figs. 2, 5, and 8). It is interesting to remark that all the illustrative examples on the plasmonic wave spectrum presented here reaffirm that the dispersive metamaterial components in the composite enable the structure to support the simultaneous existence of the TM and the TE modes. This effect is solely attributed to the negative-index metamaterials and is otherwise impossible.


\begin{figure}[htbp]
\includegraphics*[width=9cm,height=10cm]{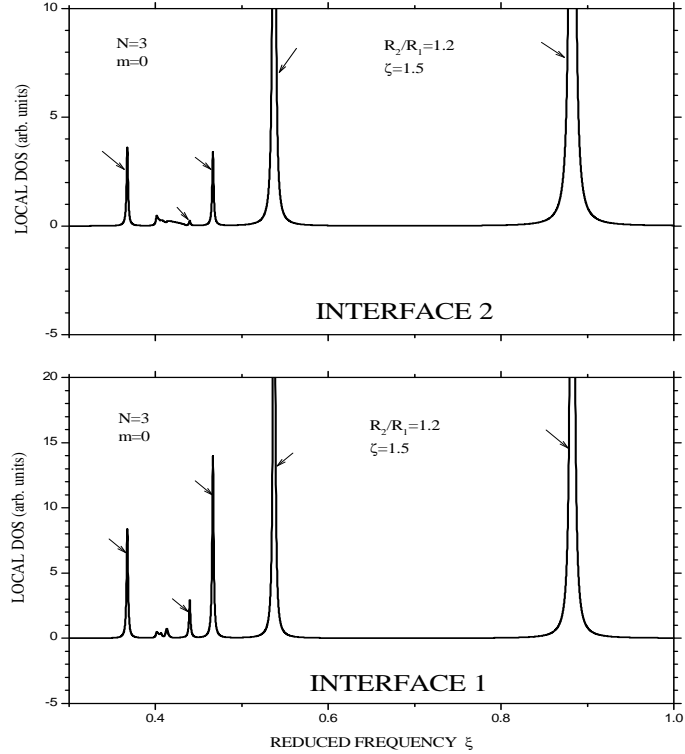}
\caption{Local density of states at the interface $R_1$ ($R_2$) in the lower (upper) panel for $m=0$ and
$\zeta=1.5$ for the system discussed in Fig. 11. We call attention to the DOS resonance peaks, indicated by the 
arrows, corresponding to the five modes in total at $\zeta=1.5$ in Fig. 11. The interface 1 (2) refers to the one specified by $R_1$ ($R_2$). The rest of the parameters used are the same as in Fig. 11.}
\label{fig12}
\end{figure}

Figure 12 shows the local density of states for the two-interface system discussed in Fig. 11 for $m=0$ and $\zeta=1.5$ for the interface 1 (2) in the lower (upper) panel. This value of $\zeta$ characterizes five propagating modes in total in Fig. 11: the lowest split (TM) mode below $\omega_0$, lower TE mode, lower (split) TM mode, upper TE mode, and the uppermost TM mode lying, respectively, at $\xi=0.3673$, $\xi=0.4396$, $\xi=0.4665$, $\xi=0.5373$, and $\xi=0.8825$. In the lower panel, the five resonance peaks (indicated by arrows) observed in the local density of states stand exactly at these frequencies. This implies considerably a very good correspondence between the (dispersion) spectrum and the LDOS at interface $R_1$. The small (unmarked) noisy peaks occurring in the vicinity of the resonance frequency $\omega_0$ are not considered to be a bonafide peaks in the LDOS. In the upper panel, we plot the LDOS for the interface 2 for the same parameters as considered for interface 1 in the lower panel. We observe five well-defined resonance peaks lying exactly at the aforementioned frequencies. That means that both interfaces share all the five resonances in the LDOS, of course with a difference of magnitude. As to the second small resonance peak in this panel, we think that interface 2 only slightly feels this resonance. Other resonance peaks in this panel are almost comparable to those in the lower panel. The rest of the remarks made with respect to Fig. 9 are also valid here.


\begin{figure}[htbp]
\includegraphics*[width=9cm,height=10cm]{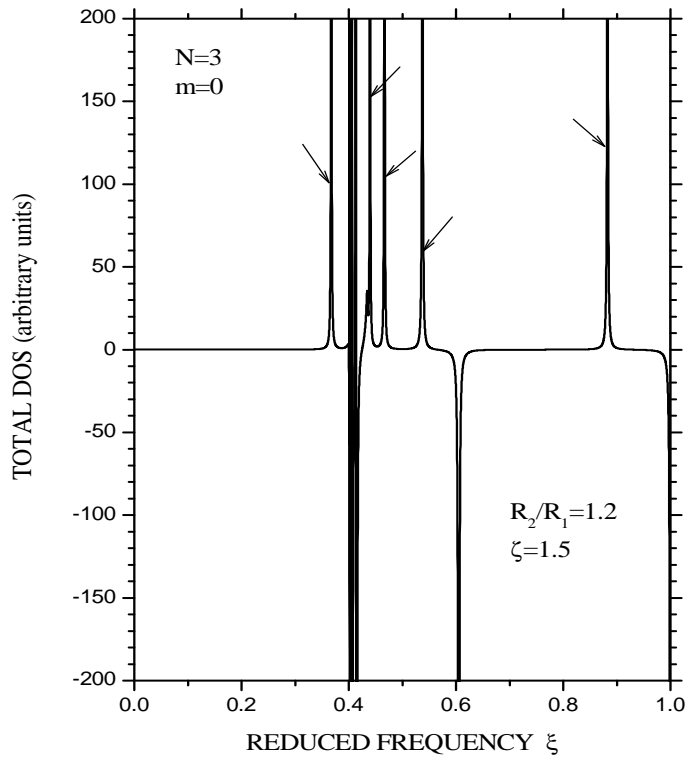}
\caption{Total density of states for $m=0$ and $\zeta=1.5$ for the system discussed in Fig. 11. We call attention to 
the DOS resonance peaks, indicated by the arrows, corresponding to the five modes in total at $\zeta=1.5$ in Fig. 11. 
The parameters used are the same as in Fig. 11. The DOS are shown in arbitrary units throughout.}
\label{fig13}
\end{figure}

Figure 13 depicts the total density of states for the same two-interface system as investigated in Figs. 11 and 12 for $m=0$ and $\zeta=1.5$. Such values of $m$ and $\zeta$ characterize five propagating modes in total in Fig. 11, covering both TM and TE modes, and lying at $\xi=0.3673$, $\xi=0.4396$, $\xi=0.4665$, $\xi=0.5373$, and $\xi=0.8825$. We observe that there are five well-defined resonance peaks in the total density of states standing exactly at the aforementioned frequencies. This leads us to infer that there is a very good correspondence between the (dispersion) spectrum and the TDOS. It has been observed that scanning other values of the propagation vector $\zeta$ for computing the total density of states yields same degree of correspondence with the spectrum.  What is more interesting in this case is the fact that the computation of TDOS (as well as LDOS) provides a much better correspondence with the modes in the spectrum even in the radiative region (toward the left of the light line) than in the previous cases. This is attributed to a relatively smoother propagation of the modes in the radiative region in the present case of a dielectric shell bounded by (identical) metamaterials. Just as before, we do not give much importance to the pile up of the states near the resonance frequency $\omega_0$ and the negative peak at $\omega_c$. While we consider their occurrence as natural, they do not bear any physical significance to the problem.

\section{CONCLUDING REMARKS}

In summary, we have investigated the plasmonic wave dispersion and the density of states in the coaxial cables in the absence of an applied magnetic field. We derived the general dispersion relations using a Green-function (or response function) theory in the framework of IRT, which has now found widespread use to study numerous excitations in various composite systems. In doing so, we not only clarify some basic notions in the use of the cylindrical geometries but also diagnose our general analytical results under special limits to reproduce some well-known results on the 2D and 1D plasmon dispersion in quantum wells and quantum wires. We have also successfully attempted to substantiate our results on plasmonic wave dispersion through the computation of the local and total density of states. While we considered the effect of retardation, the absorption was neglected throughout, except for a small imaginary part needed to be added to the frequency for the purpose of giving a width to the peaks in the DOS. We believe that the present methodology for coaxial cables will also prove to be a powerful theoretical framework for studying, for example, the intrasubband plasmons in the multi-walled carbon nanotubes.

An experimental observation of the radiative as well as non-radiative plasmonic waves in such coaxial cables would be of great interest. Such experiments could possibly involve the well known attenuated total reflection, scattering of high energy electrons, or even Raman spectroscopy. The electron energy loss spectroscopy (EELS) is already becoming known as a powerful technique for studying the electronic structure, dielectric properties, and plasmon excitations in carbon nanotubes and carbon onions, for example. Our preference for plotting the illustrative numerical results in terms of the dimensionless frequency and propagation vector leaves free an option of choosing the plasma frequency lower or higher, just as the radii of the cables.

Future dimensions worth adding to the problem remain open in this context. The issues, which need to be considered and which could give better insight into the problem, include the role of absorption, the effects of the spatial dispersion, the coupling to the optical phonons, effect of an applied electric field, and most importantly the effect of an applied magnetic field in order to study, for example, the edge magnetoplasmons in the concentric cylindrical cables, to name a few. Choosing the unidentical dielectrics and or unidentical metamaterials will only alter the asymptotic limits in the short wavelength limit.

Currently, we have been generalizing this theory to be applicable to the multicoaxial cables (where there should be no limit on the number of interfaces) and investigating the effects of an applied magnetic field in the Faraday geometry
on the plasmonic wave dispersion in such concentric cable structures and the results will be reported shortly.

\acknowledgments

One of us (M.S.K.) gratefully acknowledges the hospitality of the UFR de Physique of the University of Science and Technology of Lille 1, France, during the short visit in 2009. We sincerely thank Leonard Dobrzynski for many very fruitful discussions and communications. We also thank Jerome Vasseur for help with the software.



\begin{references}
\bibitem[1]{1} For an extensive review of electronic, optical, and transport properties of systems of
               reduced dimensionality, such as quantum wells, wires, dots, and electrically/magnetically
               modulated 2D systems, see M. S. Kushwaha, ``Plasmons and magnetoplasmons in semiconductor heterostructures", Surf. Sci. Rep. {\bf 41}, 1-416 (2001).
\bibitem[2]{2} V.G. Veselago, ``The electrodynamics of substances with simultaneously negative values of
               $\epsilon$ and $\mu$", Sov. Phys. Usp. {\bf 10}, 509-514 (1968).
\bibitem[3]{3} J.B. Pendry,  ``Negative Refraction Makes a Perfect Lens", Phys. Rev. Lett. {\bf 85}, 3966-3969
               (2000).
\bibitem[4]{4} R.A. Shelby, D.R. Smith, and S. Schultz, ``Experimental verification of a negative index of
               refraction", Science {\bf 292}, 77-79 (2001).
\bibitem[5]{5} J. Pacheco, T.M. Grzegorczyk, B.I. Wu, Y. Zhang, and J.A. Kong, ``Power Propagation
               in Homogeneous Isotropic Frequency-Dispersive Left-Handed Media", Phys. Rev. Lett.
               {\bf 89}, 257401 (2002).
\bibitem[6]{6} J. Li, L. Zhou, C.T. Chan, and P. Sheng,  ``Photonic Band Gap from a Stack of Positive
               and Negative Index Materials", Phys. Rev. Lett. {\bf 90}, 083901 (2003).
\bibitem[7]{7} D. Bria, B. djafari-Rouhani, A. Akjouj, L. Dobrzynski, J. P. Vigneron, E.H. EL Boudouti,
               and A. Nougaoui, ``Band structure and omnidirectional photonic band gap in lamellar
               structures with left-handed materials", Phys. Rev. B {\bf 69}, 066613 (2004).
\bibitem[8]{8} I.V. Shadrivov, A.A. Sukhorukov, and Y.S. Kivshar, ``Complete Band Gaps in One-Dimensional
               Left-Handed Periodic Structures", Phys. Rev. Lett. {\bf 95}, 193903 (2005).
\bibitem[9]{9} N.C. Panoiu, R.M. Osgood, S. Zhang, and S.R.J. Brueck, ``Zero-n bandgap in photonic crystal
               superlattices", J. Opt. Soc. Am. B {\bf 23}, 506-513 (2006).
\bibitem[10]{10} B. Wood and J.B. Pendry, ``Metamaterials at zero frequency", J. Phys. : Condens. Matter
                 {\bf 19}, 076208 (2007).
\bibitem[11]{11} W.J. Hsueh, C.T. Chen, and C.H. Chen, ``Omnidirectional band gap in Fibonacci photonic
                 crystals with metamaterials using a band-edge formalism", Phys. Rev. A {\bf 78}, 013836 (2008).
\bibitem[12]{12} Y. Wu and Z.Q. Zhang, ``Dispersion relations and their symmetry properties of electromagnetic and
                 elastic metamaterials in two dimensions", Phys. Rev. B {\bf 79}, 195111 (2009).
\bibitem[13]{13} R. Ruppin, ``Surface polaritons of a left-handed medium", Phys. Lett. A {\bf 277}, 61-64 (2000).
\bibitem[14]{14} I.V. Shadrivov, A.A. Shukhorukov, and Y.S. Kivshar, ``Guided modes in negative-refractive-index
                 waveguides", Phys. Rev. E {\bf 67}, 057602 (2003).
\bibitem[15]{15} S.A. Darmanyan, M. Neviere, and A.A. Zakhidov, ``Surface modes at the interface of
                 conventional and left-handed media", Opt. Commun. {\bf 225}, 233-240 (2003).
\bibitem[16]{16} H. Cory and A. Barger, ``Surface-wave propagation along a metamaterial slab", Microw. Opt.
                 Technol. Lett. {\bf 38}, 392-395 (2003).
\bibitem[17]{17} H. Cory and C. Zach, ``Wave propagation in metamaterial multi-layered structures", Microw. Opt.
                 Technol. Lett. {\bf 40}, 460-465 (2004).
\bibitem[18]{18} Y. He, Z. Cao, and Q. Shen, ,``Guided optical modes in asymmetric left-handed waveguides"
                 Opt. Commun. {\bf 245}, 125-135 (2005).
\bibitem[19]{19} L.G. Wang, H. Chen, and S.Y. Zhu, ``Wave propagation inside one-dimensional photonic crystals
                 with single-negative materials", Phys. Lett. A {\bf 350}, 410-415 (2006).
\bibitem[20]{20} Y. Huang, Y. Feng, and T. Jiang, ``Electromagnetic cloaking by layered structure of homogeneous
                 isotropic materials", Opt. Express {\bf 15}, 11133-11141 (2007).
\bibitem[21]{21} Y. Fang and S. He, ``Transparent structure consisting of metamaterial layers and matching
                 layers", Phys. Rev. A {\bf 78}, 023813 (2008).
\bibitem[22]{22} F. Tao, H.F. Zhang, X.H. Yang, and D. Cao, ``Surface plasmon polaritons of the metamaterial
                 four-layered structures", J. Opt. Soc. Am. B {\bf 26}, 50-59 (2009).
\bibitem[23]{23} V. Kuzmiak and A.A. Maradudin, ``Scattering properties of a cylinder fabricated from a left-handed
                 material ", Phys. Rev. B {\bf 66}, 045116 (2002).
\bibitem[24]{24} N.C. Paniou and R.M. Osgood, ``Numerical investigation of negative refractive index metamaterials
                 at infrared and optical frequencies", Opt. Commun. {\bf 223}, 331-337 (2003).
\bibitem[25]{25} R. Ruppin, ``Surface polaritons and extinction properties of a left-handed material cylinder",
                 J. Phys.: Condens. Matter {\bf 16}, 5991-5998 (2004).
\bibitem[26]{26} S. Ancey, Y. Decanini, A. Folacci, and P. Gabrielli, ``Surface polaritons on left-handed cylinders:
                 A complex angular momentum analysis", Phys. Rev. B {\bf 72}, 085458 (2005).
\bibitem[27]{27} H. Cory and T. Blum, ``Surface-wave propagation along a metamaterial cylindrical guide", Microw.
                 Opt. Technol. Lett. {\bf 44}, 31-35 (2005).
\bibitem[28]{28} K.Y. Kim, J.H. Li, Y.K. Cho, and H.S. Tae, ``Electromagnetic wave propagation through doubly
                 dispersive subwavelength metamaterial hole", Opt. Express {\bf 13}, 3653-3665 (2005).
\bibitem[29]{29} S. Arslanagic, R.W. Ziolkowski and O. Breinbjerg, ``Excitation of an electrically small
                 metamaterial-coated cylinder by an arbitrarily located line source", Microw. Opt. Technol. Lett.
                 {\bf 48}, 2598-2606 (2006).
\bibitem[30]{30} E. Irci and V.K. Erturk, ``Achieving transparency and maximizing scattering with metamaterial-coated
                 conducting cylinders", Phys. Rev. E {\bf 76}, 056603 (2007).
\bibitem[31]{31} K.Y. Kim, ``Fundamental guided electromagnetic dispersion characteristics in lossless dispersive
                 metamaterial clad circular air-hole waveguides", J. Opt. A: Pure Appl. Opt. {\bf 9}, 1062-1069 (2007).
\bibitem[32]{32} S. Ahmed and Q.A. Naqvi, ``Electromagnetic scattering from a perfect electromagnetic conductor
                 circular cylinder coated with a metamaterial having negative permittivity and/or permeability", Opt. Commun. {\bf 281}, 5664-5670 (2008).
\bibitem[33]{33} H.Y. She, L.W. Li, O.J.F. Martin, and J.R. Mosig, ``Surface polaritons of small coated cylinders
                 illuminated by normal incident TM and TE plane waves", Opt. Express {\bf 16}, 1007-1019 (2008).
\bibitem[34]{34} C. Garcia-Meca, R. Ortuno, F.J. Rodriguez, J. Marti, and A. Martinez, ``Negative refractive index
                 metamaterials aided by extraordinary optical transmission", Opt. Express {\bf 17}, 6026-6031 (2009).
\bibitem[35]{35} K.L. Tsakmakidis, A.D. Boardman, and O. Hess, ``Trapped rainbow storage of light in metamaterials",
                 Nature {\bf 450}, 397-401 (2007).
\bibitem[36]{36} J. Lee and J.B. Pendry, ``Hiding under the Carpet: A New Strategy for Cloaking", Phys. Rev. Lett.
                 {\bf 101}, 203901 (2008).
\bibitem[37]{37} M.G. Silveirinha, ``Anomalous Refraction of Light Colors by a Metamaterial Prism", Phys. Rev. Lett.
                {\bf 102}, 193903 (2009).
\bibitem[38]{38} J.B. Pendry and D.R. Smith, ``Reversing light with negative refraction", Physics Today {\bf 57}(6),
                 37-44 (2004).
\bibitem[39]{39} A.D. Boardman, N. King, and L. Velasco, ``Negative Refraction in Perspective", Electromagnetics
                 {\bf 25}, 365-389 (2005).
\bibitem[40]{40} T.W. Ebbesen, H.J. Lezec, H. Ghaemi, T. Thio, and P.A. Wolf, ``Extraordinary optical transmission
                through sub-wavelength hole arrays", Nature {\bf 391}, 667-669 (1998).
\bibitem[41]{41} J.A. Porto, F.J. Garcia-Vidal, and J.B. Pendry, ``Transmission Resonances on Metallic Gratings
                 with Very Narrow Slits", Phys. Rev. Lett. {\bf 83}, 2845-2848 (1999).
\bibitem[42]{42} L. Martin-Moreno, F.J. Garcia-Vidal, H.J. Lezec, K.M. Pellerin, T.Thio, and J.B. Pendry, and
                 T.W. Ebbesen, ``Theory of Extraordinary Optical Transmission through Subwavelength Hole Arrays",
                 Phys. Rev. Lett. {\bf 86}, 1114-1117 (2001).
\bibitem[43]{43} Y.Takakura, ``Optical Resonance in a Narrow Slit in a Thick Metallic Screen", Phys. Rev. Lett.
                 {\bf 86}, 5601-5603 (2001).
\bibitem[44]{44} J.B. Pendry, L. Martin-Moreno, and F.J. Garcia-Vidal, ``Mimicking Surface Plasmons with Structured
                 Surfaces", Science {\bf 305}, 847-848 (2004.)
\bibitem[45]{45} D.R. Smith, D.C. Vier, W. Padilla, S.C. Nemat-Nasser, and S. Schultz, ``Loop-wire medium for
                 investigating plasmons at microwave frequencies", Appl. Phys. Lett. {\bf 75}, 1425-1427 (1999).
\bibitem[46]{46} F. Yang and J.R. Sambles, ``Resonant Transmission of Microwaves through a Narrow Metallic Slit",
                 Phys. Rev. Lett. {\bf 89}, 063901 (2002).
\bibitem[47]{47} J.R. Suckling, A.P. Hibbins, M.J. Lockyear, T.W. Preist, J.R. Sambles, and C.R. Lawrence,
                 ``Finite Conductance Governs the Resonance Transmission of Thin Metal Slits at Microwave
                 Frequencies", Phys. Rev. Lett. {\bf 92}, 147401 (2004).
\bibitem[48]{48} S.A. Maier, S.R. Andrews, L. Martin-Moreno, and F.J. Garcia-Vidal, ``Terahertz Surface
                 Plasmon-Polariton Propagation and Focusing on Periodically Corrugated Metal Wires", Phys. Rev.
                 Lett. {\bf 97}, 176805 (2006).
\bibitem[49]{49} Z. Chen, I.R. Hooper, and J.R. Sambles, ``Strongly coupled surface plasmons on thin shallow
                 metallic gratings", Phys. Rev. B {\bf 77}, 161405 (2008).
\bibitem[50]{50} A.P. Hibbins, M.J. Lockyear, I.R. Hooper, and J.R. Sambles, ``Waveguide Arrays as Plasmonic
                 Metamaterials: Transmission below Cutoff", Phys. Rev. Lett. {\bf 96}, 073904 (2006).
\bibitem[51]{51} A.P. Hibbins, M.J. Lockyear, and J.R. Sambles, ``Coupled surface-plasmon-like modes between
                 metamaterial", Phys. Rev. B {\bf 76}, 165431 (2007).
\bibitem[52]{52} M.J. Lockyear, A.P. Hibbins, and J.R. Sambles, ``Microwave Surface-Plasmon-Like Modes on Thin
                 Metamaterials", Phys. Rev. Lett. {\bf 102}, 073901 (2009).
\bibitem[53]{53} L. Dobrzynski, ``Interface response theory of discrete composite systems", Surf. Sci. Rep.
                 {\bf 6}, 119-157 (1986); ``Response theory of interfaces, superlattices, and composite materials",
                 Surf. Sci. {\bf 300}, 1008-1021 (1994).
\bibitem[54]{54} L. Dobrzynski and H. Puszkarski, ``Eigenvectors of composite systems. I. General theory",
                 J. Phys.: Condens. Matter {\bf 1}, 1239-1245 (1989).
\bibitem[55]{55} M.S. Kushwaha and B. Djafari-Rouhani, ``Theory of magnetoplasmons in semiconductor superlattices
                 in the Voigt geometry: A Green-function approach", Phys. Rev. B {\bf 43}, 9021-9032 (1991).
\bibitem[56]{56} B. Djafari-Rouhani and L. Dobrzynski, ``Acoustic resonances of adsorbed wires and channels", J.
                 Phys.: Condens. Matter Matter {\bf 5}, 8177-8194 (1993).
\bibitem[57]{57} P.M. Morse and H. Feshbach, {\it Methods of Theoretical Physics} (McGraw-Hill, New
                 York, 1953), Vol. I, Chapt. 7.
\bibitem[58]{58} M. Abramowitz and I.A. Stegun, {\it Handbook of Mathematical Functions}
                 (Dover, New York, 1972).
\bibitem[59]{59} See, for example, R. Ruppin, in: A.D. Boardman, ed., {\it Electromagnetic Surface
                 Modes} (Wiley, New York, 1982), p. 345-398.
\bibitem[60]{60} J. Wang and J.P. Leburton, ``Plasmon dispersion relation of a quasi-one-dimensional electron
                 gas", Phys. Rev. B {\bf 41}, 7846-7849 (1990).
\bibitem[61]{61} Q.P. Li and S. Das Sarma, ``Elementary excitation spectrum of one-dimensional electron systems
                 in confined semiconductor structures: Zero magnetic field", Phys. Rev. B {\bf 43}, 11768-11786
                 (1991).

\end{references}
\end{document}